\documentclass[fleqn,usenatbib]{mnras}

\usepackage{newtxtext,newtxmath}

\usepackage[T1]{fontenc}
\usepackage{ae,aecompl}
\usepackage{color}

\usepackage{graphicx}	
\usepackage{amsmath}	
\usepackage{amssymb}	

\title{Forecasts for Warm Dark Matter from photometric galaxy surveys}

\author[J. Martins,
R. Rosenfeld and F. Sobreira]{
J\'essica S. Martins$^{1,2}$\thanks{E-mail: jessica@ift.unesp.br},
Rogerio Rosenfeld$^{1,2}$\thanks{E-mail: rosenfel@ift.unesp.br},
Flavia Sobreira$^{2,3}$\thanks{E-mail: sobreira@ifi.unicamp.br}
\\
$^{1}$Instituto de F\'isica Te\'orica, UNESP \& ICTP-SAIFR, Rua Dr. Bento Teobaldo Ferraz, 271, 01140-070 - S\~ao Paulo, Brazil\\
$^{2}$Laborat\'orio Interinstitucional de e-Astronomia-LineA, Rua Gal. Jos\'e Cristino, 77, 20921-400 - Rio de Janeiro, Brazil \\
$^{3}$Instituto de F\'{\i}sica Gleb Wataghin, Universidade Estadual de Campinas, 13083-859, Campinas, SP, Brazil
}

\date{Accepted XXX. Received YYY; in original form ZZZ}

\pubyear{2018}

\begin{document}
\label{firstpage}
\pagerange{\pageref{firstpage}--\pageref{lastpage}}
\maketitle

\begin{abstract}
We present a Fisher matrix forecast for the sensitivity on the mass of a thermal warm dark matter (WDM) particle from current (DES-like) and future (LSST-like) photometric galaxy surveys using the galaxy angular power spectrum. We model the nonlinear clustering using a modified Halo Model proposed to account for WDM effects. We estimate that from this observable alone a lower bound of $m_{\text{wdm}}>647$\,eV ($m_{\text{wdm}}>126$\,eV) for the LSST (DES) case could be obtained.
\end{abstract}

\begin{keywords}
large-scale structure of Universe -- dark matter -- surveys
\end{keywords}



\section{Introduction}

The $\Lambda$CDM model, with a cosmological constant ($\Lambda$) and cold dark matter (CDM) contributing approximately $70 \%$ and $25 \%$ respectively to the energy density budget, is the best cosmological description of our universe we have to-date. This conclusion comes from a variety of observations from different probes at different epochs. A recent example is the analysis of the first year of data of the Dark Energy Survey using probes from galaxy clustering and weak lensing simultaneously to show the consistency of this model even when combined with data from the cosmic microwave background \citep{2017arXiv170801530D}.  

However, it is fair to say that the nature of dark matter is still not settled. In fact, some tensions have been found when comparing small scales (few Mpc down to Kpc) observations with CDM-only numerical simulations.
These tensions can be described by three ``problems": the core-cusp problem, where 
the inner density profile of a CDM halo in a simulation with dark matter only has a cuspy density profile close to the centre of the halo whereas the measured density of a galaxies has a core profile for small radii \citep{diemand,salucci,swaters,de2010core,oh2011dark};
the missing satellite problem, which arises because one observes less satellite galaxies of our galaxy and M31 than subhaloes predicted in CDM simulations \citep{klypin1999missing,moore,bullock2013notes};
and the  `\textit{too-big-to-fail}' problem, where haloes that are massive enough to form dwarf galaxies in simulations are not actually found in observations, that is, the observed dwarf galaxies are less massive then predicted \citep{tikhonov2009emptiness,peebles2010nearby,garrison2014too}.

While several groups try to explain these tensions through astrophysical processes such as adding baryons in simulations \citep{2016arXiv160706479F}, there is also the possibility of changing the nature of dark matter to obtain a better agreement with observations. For instance, assuming that dark matter is warm instead of cold could in principle ammeliorate these tensions. A recent comparison between the WDM and baryonic effects in the context of the too-big-to-fail problem can be found in \cite{2017MNRAS.468.2836L}. 

Regardless of these issues with simulations, the nature of the dark matter is a fundamental question for particle physics and should be investigated using any available probe. The aim of this paper is to study the possibility of using the angular power spectrum of photometric galaxy surveys to answer the question of whether dark matter is cold or warm.

Warm dark matter behaves very similarly to CDM at large scales but in the early universe it decouples while still mildly relativistic. This gives a thermal velocity to the dark matter particles and consequently a non-negligible free-streaming scale below which perturbations are smoothed out.
The tightest constraint on the warm dark matter particle mass comes from Lyman~$\alpha$ (Ly~$\alpha$) forest from absorption lines of distant quasars in the intergalactic medium at high redshifts and it reaches a lower bound of $m_{\text{wdm}}\geq 5.3$\,keV (at $2 \sigma$ CL) if warm dark matter is assumed to be a thermal relic \citep{lyalphamx}.

In this work we study the sensitivity to the mass of a thermal warm dark matter particle using a Fisher matrix approach considering the galaxy angular power spectrum in photometric surveys as an observable. We will use as examples the Dark Energy Survey (DES) and a Large Synoptic Survey Telescope (LSST)-like surveys. The ongoing DES\footnote{www.darkenergysurvey.org} project is a wide area ($\sim 5,000$ deg$^2$) and relatively deep ($z\sim$ 1.4) photometric map of the southern sky and among its goals is the determination of the cosmological parameters using the distribution of galaxies, weak gravitational lensing, cluster number counts and type Ia supernovae. The LSST\footnote{www.lsst.org} is intended to be the largest galaxy survey ever made mapping 30,000 deg$^{2}$ of the visible sky for $z\leq 3$ and will be able to perform a variety of studies, including the investigation on the nature of dark energy and dark matter.

One particular challenge we face comes from the fact that the modifications due to warm dark matter in the power spectrum arise at small, nonlinear scales. We adopt a halo model prescription to estimate the power spectrum at these scales. The halo model provides a flexible tool to model nonlinear effects for a given input cosmology. It may be used as a less computationally intensive, albeit less accurate, alternative to full-fledged simulations in exploratory studies such as the present one. However, we will also show results using a numerical fit to warm dark matter simulations.

We find our forecasts to be less restrictive than the Ly~$\alpha$ constraints, but these bounds should be pursued anyways in combination with other probes.

This paper is organized as follows: in Section \ref{sec:freewdm} we review the free-streaming mechanism for a thermal relic in the linear regime, the halo model description of the power spectrum in the non-linear regime and a its extension to describe WDM; in Section \ref{sec:Fisher} we develop the Fisher matrix analysis for a DES-like and a LSST-like surveys to obtain the sensitivities on the dark matter mass and in Section \ref{sec:discussion} we discuss our results and present our conclusions.

\section{Warm dark matter and structure formation}\label{sec:freewdm}

\subsection{Linear regime}

Warm dark matter, being lighter than its CDM counterpart, remains relativistic for a longer period during the radiation dominated era and also retains some thermal velocity at decoupling. This gives enough time for warm dark matter particles to diffuse out of perturbations after their decoupling. The effect of this scenario at late times is a suppression on structure formation below a certain scale related to the free-streaming length of the particles, which depends on their mass.

A simple way to estimate the free-streaming length is by computing the comoving length scale that a particle can travel until matter-radiation equality ($t_{\text{eq}}$) \citep{kolb}:

	\begin{align}
	\lambda_{\text{fs}} = \int_{0}^{t_{\text{eq}}}\frac{v(t)dt}{a(t)} \approx \int_{0}^{t_{\text{NR}}}\frac{cdt}{a(t)} + \int_{t_{\text{NR}}}^{t_{\text{eq}}}\frac{v(t)dt}{a(t)},
	\end{align}
where $t_{\text{NR}}$ is the time when WDM particles become non-relativistic. 


For WDM made of a two-component fermion, the free-streaming length can be written as:
	\begin{align}
	\lambda_{\text{fs}}\approx 0.4 \left(\frac{m_{\text{wdm}}}{\text{keV}} \right)^{-4/3} \left(\frac{\Omega_{\text{wdm}}h^{2}}{0.135} \right)^{1/3} [h^{-1}\text{Mpc}],
	\end{align}
where $m_{\text{wdm}}$ and $\Omega_{\text{wdm}}$ are the mass and density parameter of the warm dark matter particle, respectively. In this work we will assume that all dark matter in the universe is warm, when calculating constraints on its mass.

The free-streaming scale can be qualitatively illuminating but to obtain a more accurate scenario of the WDM physics first we need the transfer function for this type of dark matter. We work here with the \citet{bode} fitting formula from Boltzmann code calculations, with revisited parameters \citep{vielwdm}:
	\begin{align}
	T_{\text{wdm}}(k) = \left[ \frac{P^{\text{lin}}_{\text{wdm}}(k)}{P^{\text{lin}}_{\text{cdm}}(k)} \right] ^{1/2} = \left[1 + \left(\alpha k \right)^{2\mu}  \right]^{-5/\mu}, \label{eq:transfer}
	\end{align}
where $\mu = 1.12$ and,
	\begin{align}
	\alpha = 0.049 \left( \frac{m_{\text{wdm}}}{\text{keV}}\right)^{-1.11} \left( \frac{\Omega_{\text{wdm}}}{0.25}\right)^{0.11}\left( \frac{h}{0.7} \right)^{1.22} [h^{-1}\text{Mpc}].
	\end{align}

	\begin{figure}
	\includegraphics[width=\columnwidth]{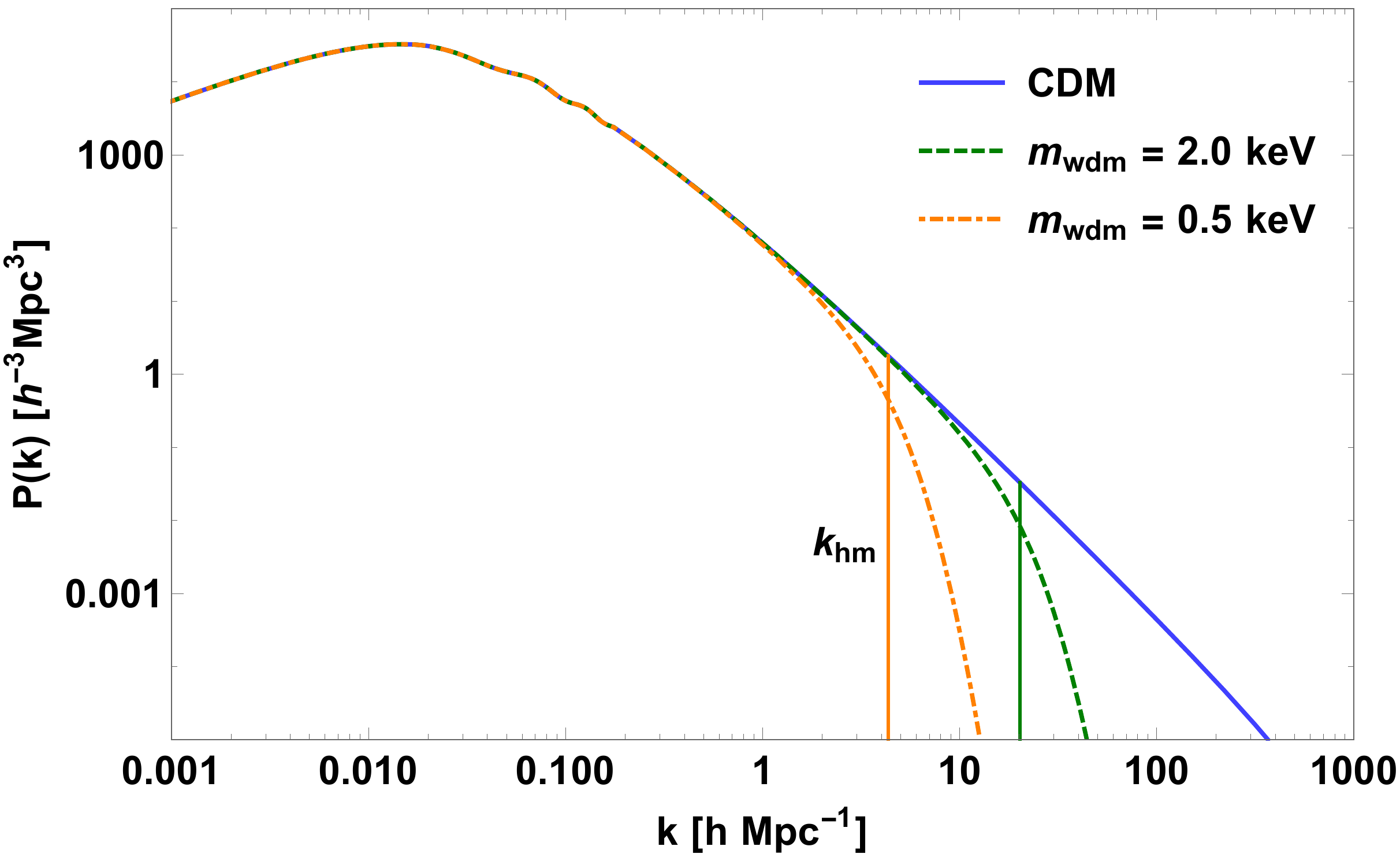}
	\caption{Linear matter power spectrum at $z=0$ for WDM particle masses of $m_{\text{wdm}}=2.0$\,keV (dashed) and $m_{\text{wdm}}=0.5$\,keV (dot-dashed), together with the linear power spectrum for CDM. The vertical lines indicate the half-mode scale for each mass.}
	\label{fig:Plinwdm}
	\end{figure}

In Fig.~\ref{fig:Plinwdm} we show the linear power spectrum for warm dark matter. We see that, as expected, the lighter the WDM particle is, the more it will suppress the formation of structure, since it stays relativistic for a longer time.

The characteristic length-scale $\alpha$ in the parametrization of the transfer function is closely related to the free-streaming scale $\lambda_{\text{fs}}$, and we will define $\lambda^{\text{eff}}_{\text{fs}}\equiv \alpha$ as an effective free-streaming scale. This scale can be used to define a free-streaming mass scale given by:
	\begin{align}
	M_{\text{fs}}(z) = \frac{4\pi}{3}\bar{\rho}(z)\left(\frac{\lambda^{\text{eff}}_{\text{fs}}}{2} \right)^{3},
	\end{align}
where $\bar{\rho}(z)$ is the mean density of the universe at a given redshift. This free-streaming mass defines the mass scale where the suppression of structure formation occurs. Below this scale the initial density perturbation are mostly erased.

Another useful length scale introduced in the literature is the half-mode scale $\lambda_{\text{hm}}$, which corresponds to the length scale at which the amplitude of the WDM transfer function is suppressed to 1/2 relative to CDM. From equation~(\ref{eq:transfer}) we get for the half-mode scale:
	\begin{align}
	\lambda_{\text{hm}} = 2\pi \lambda^{\text{eff}}_{\text{fs}}\left(2^{\mu/5}-1 \right)^{-1/2\mu} \approx 14 \lambda^{\text{eff}}_{\text{fs}} .
	\end{align}

This scale is shown as the vertical lines ($k_{\text{hm}}=2\pi/\lambda_{\text{hm}}$) of Fig.~\ref{fig:Plinwdm} for WDM particles of masses $m_{\text{wdm}}=2.0$\,keV and $m_{\text{wdm}}=0.5$\,keV. As expected this scale is larger for smaller masses. 

The half-mode length scale leads to another mass scale, called the half-mode mass scale:
	\begin{align}
	M_{\text{hm}}(z) = \frac{4\pi}{3}\bar{\rho}(z)\left(\frac{\lambda_{\text{hm}}}{2} \right)^{3} \approx 2.7\times 10^{3}M_{\text{fs}}(z).\label{eq:Mhm}
	\end{align}

The half-mode mass scale was found from numerical simulations to be the relevant one where WDM physics first affect the properties of structure formation \citep{colin2008structure,menci2012galaxy}. 


\subsection{Non-linear regime: the Standard Halo Model}

At low redshifts the non-linear effects of gravity become relevant and modify the predictions of the linear theory. In fact, non-linearities tend to increase the power spectrum at small scales due to gravitational clumping.
Unfortunately, it was found from numerical simulations that not much information is retained from the the linear power spectrum with a small-scale suppression after the nonlinear growth of structure \citep{2017arXiv171202742L}. This effect obviously reduces the sensitivity to the mass of the WDM particle which causes the suppression in the linear power spectrum in the first place.


Therefore, in order to get meaningful results, we have to take into account the non-linear effects of gravity. One approach is by running simulations such as N-body or hydrodynamical simulations, but these are very costly and time-consuming as they require large computers and need to be repeated for each different cosmology. Another strategy available is to make use of semi-analytical models such as the halo model, which gives somewhat accurate results when compared to simulations \citep{massara2014halo,white2001redshift} and enables qualitative insights about structure formation at non-linear scales. We briefly recap the basics of the standard halo model below since we will modify it slightly in the WDM case.

The basic assumption of the halo model is that all the matter in the universe is inside virialized dark matter objects called halos \citep{sheth,zentner2005halo}. The halos form from overdensities in the matter density field that are greater than the critical density $\delta_{\text{sc}}\approx 1.686$, a parameter derived from the spherical collapse (sc) model that quantifies the critical overdensity necessary for structures to collapse. The halo model is completely determined by three ingredients: the halo mass function, the halo density profile and the halo bias. 

The halo mass function accounts for the number density of halos for a given mass and redshift, and can be estimated by the number of overdensities above the critical density at some redshift. If one assumes that the initial fluctuations of the density field are Gaussian one gets for the mass function \citep{bardeen1986statistics},
	\begin{align}
	\frac{dn}{dM}(M,z)dM = \frac{\bar{\rho}(z)}{M}\nu f(\nu)\frac{d\nu}{\nu},
	\end{align}
where $\nu(M,z) = \delta_{sc}^{2}/\sigma(M,z)^{2}$ and $\sigma(M)$ is the variance of the matter density field given by
	\begin{align}
	\sigma(M,z)^{2} = \frac{1}{2\pi^{2}}\int P(k)D(z)^{2}\tilde{W}_{\text{T}}(kR)^{2}k^{2}dk,\qquad M=\frac{4\pi R^{3}}{3}\bar{\rho}, \label{variancia}
	\end{align}
where $D(z)$ is the growth function and $\tilde{W}_{\text{T}}(kR)$ is the Fourier transform of the top-hat window function in real space:
	\begin{align}
	W_{\text{T}}(x) = \frac{3(\sin(x)-x\cos(x))}{x^{3}}.
	\end{align}

Finally, the function $f(\nu)$ is,
	\begin{align}
	\nu f(\nu)=\sqrt{\frac{\nu}{2\pi}}\exp(-\nu/2).
	\end{align}

This is the Press--Schechter mass function \citep{pressschechter} and it gives a reasonable description of number density of halos when compared to simulations. Inspired by the Press \& Schechter formalism, many mass functions \citep{shethellipsoidal,tinker2008toward} were obtained later from numerical simulations of cold dark matter, such as the Jenkins mass function \citep{pssimulations}, which will be used in this work
	\begin{align}
	\nu f(\nu) = a_{1}\exp\left( -\left|\log\left( \frac{ \sqrt{\nu}}{\delta_{\text{sc}}}\right) +a_{2}\right|^{a_{3}}\right)\label{eq:fnu} , 
	\end{align}
where $a_{1}\approx 0.315$, $a_{2}\approx 0.61$ and $a_{3}\approx 3.8$. We also impose the normalisation condition,
	\begin{align}
	\int f(\nu)d\nu = 1,\label{eq:norm}
	\end{align}
where the integration is taken over $\nu_{\text{min}}=\nu(10^{3}~\text{M}_{\sun})$ and $\nu_{\text{max}}=\nu(10^{16}~\text{M}_{\sun})$, which guarantees that all matter in the universe is inside haloes.

The halo density profile describes how dark matter is distributed inside a halo. Assuming spherical haloes, functions of the form:
	\begin{align}
		\rho(r,M) = \frac{\rho_{\text{s}}}{\left(r/r_{\text{s}} \right)^{\alpha}\left(1+r/r_{\text{s}}\right)^{\beta}},
	\end{align}
have been widely studied in numerical simulations, where $\rho_{\text{s}}$ is the amplitude of the density profile and $r_{\text{s}}$ defines the scale of the halo radius. For $(\alpha,\beta) = (1,3)$ we have the Hernquist \citep{hernquist} profile and for $(\alpha,\beta) = (1,2)$ we have the Navarro--Frenk--White (NFW) \citep{nfw,navarro} density profile, which will be used throughout this work. 

In the case of NFW profile,
	\begin{align}
	\rho_{\text{s}} = \frac{180\bar{\rho}_{\text{z}}}{3}\frac{c^{3}}{\ln(1+c)-c/(1+c)}, \qquad 	c = \frac{11}{1+z}\left(\frac{M}{M_{\ast}} \right)^{-0.13} \label{eq:nfw},
	\end{align}
where $c\equiv r_{\text{vir}}/r_{\text{s}}$ is the concentration parameter, $r_{\text{vir}}$ is the virial radius of the halo and $M_{\ast}$ is chosen such that $\nu(M_{\ast},z) = 1$. The $180\bar{\rho}_{\text{z}}$ factor is the average density of halos and also comes from the spherical collapse model, where $\bar{\rho}_{\text{z}}$ is the average density of the universe when the halo collapsed. 

We will use below the normalized Fourier transform of the density profile,
	\begin{align}
	u(k|M) = \int_{0}^{r_{\text{vir}}} dr 4\pi r^{2} \frac{\sin(kr)}{kr}\frac{\rho(r,M)}{M}. 
	\end{align} 
	
As dark matter haloes are biased tracers of the real distribution of dark matter we also need the halo bias \citep{sheth1999large,tinker2010large,mo1996analytic}. One of the most accurate halo bias is the Tinker bias \citep{tinker2010large},
	\begin{align}
	b(\nu) = 1 + \frac{1}{\delta_{\text{sc}}}\left[q\nu + s\left( q\nu \right)^{1-t}-\frac{1}{\sqrt{q}} \frac{1 }{1+s(1-t)(1-t/2)(q\nu)^{-t}}\right] \label{eq:bias}, 
	\end{align}
where $q=0.707,\, s=0.35$ and $t=0.8$.

One of the key features of the halo model is that we can separate the physics into two different regimes: linear and non-linear scales. At small scales we consider the interaction of elements of matter inside the same halo to build the statistics and we call this term of the statistics 1-halo term. Similarly, considering the interaction of matter in different halos we describe large scales and we build the 2-halo term of the statistics. 
Therefore, in the halo model the total 3D power spectrum can be written as
	\begin{align}
	P_{\text{HM}}(k,z) = P_{\text{1h}}(k,z) + P_{\text{2h}}(k,z),
	\end{align}
with
	\begin{align}
	P_{\text{1h}}(k,z) = &\frac{1}{\bar{\rho}^{2}}\int dM M^{2}\frac{dn}{dM}(M,z) \,|u(k|M)|^{2} \label{p1h}\\ 
	P_{\text{2h}}(k,z) = &\frac{1}{\bar{\rho}^{2}}\left( \int dM M \frac{dn}{dM}(M,z)\, u(k|M)b(M,z)\right)^{2}D(z)^{2}P_{\text{lin}}(k), \label{p2h}
	\end{align}
where we have used that in the linear regime $P_{\text{hh}}(k,z|M_{1},M_{2}) \approx b_{1}(M_{1})b_{2}(M_{2}) P_{\text{lin}}(k,z)$, with $P_{\text{lin}}(k,z)$ the linear power spectrum and $P_{\text{lin}}(k,z)=D(z)^{2}P_{\text{lin}}(k,z=0)$.  

Notice that the linear power spectrum does not enter directly in the 1-halo term. 

\subsection{Halo model modifications for WDM}

The halo model has proven to be a useful tool to study cosmology at non-linear scales but it was constructed and calibrated for cold dark matter, as the halo mass function, density profile and halo bias all come from N-body simulation of CDM. In order to use the halo model for warm dark matter one needs to make some modifications. 

There are several proposals to modify the halo model for WDM in the literature, e.g. \citet{smithtesting}, \citet{dunstan2011halo}, \citet{fuzzy}, \citet{schneider2012} and \citet{schneider}. In this work we are going to adopt a recent proposal by Schneider \citep{schneider}. 

First, one assumes the window function to be a 3D spherical top-hat in Fourier space (so-called sharp-k window) instead of a top-hat in real space. The motivation to do so comes from equation~(\ref{variancia}). If we have a linear power spectrum that decreases more rapidly than $k^{-3}$ for large $k$, which is the case for warm dark matter, we loose the sensitivity of the variance over the power spectrum at non-linear scales with a top-hat window function. As a consequence, the halo model would not account for the suppression of the power at small scales. But with a sharp-k window function $W_{\text{sk}}(kr) = \Theta (1-kR)$ equation~(\ref{variancia}) becomes,
	\begin{align}
	\sigma_{\text{sk}}(R)^{2} = \frac{1}{2\pi^{2}}\int_{0}^{1/R} P^{\text{lin}}_{\text{wdm}}(k)k^{2}dk, \label{varianciask}
	\end{align}   
and now we have a variance that fully captures the WDM suppression. As the relation between radius and mass is not well defined for a sharp-k window function it is useful to impose:
	\begin{align}
	M = \frac{4\pi}{3}\bar{\rho}(bR)^{3},
	\end{align}
where $b=2.5$ is fitted from simulations \citep{benson2012dark}.

The concentration parameter is also modified using a generalization  of the CDM case:
	\begin{align}
	c_{\text{wdm}}(M) = c_{\text{cdm}}(M)\left(1 + \gamma_{1}\frac{M_{\text{hm}}(z)}{M} \right)^{-\gamma_{2}}\label{eq:cm},
	\end{align}
where the parameters $\gamma_{1}=15$ and $\gamma_{2}=0.3$ are adjusted from N-body simulations \citep{schneider2012} and $M_{\text{hm}}(z)$ is defined in equation (\ref{eq:Mhm}). We use this concentration parameter for WDM in the NFW density profile, keeping its functional form the same as for the CDM case.

We used the same functional form of the Tinker halo bias $b(\nu)$ in equation (\ref{eq:bias}) and of the Jenkins mass function in equation (\ref{eq:fnu}), but as $\nu$ is different for different $m_{wdm}$ through the variance, these quantities will also depend on the mass of WDM.

\begin{figure}
	\includegraphics[width=\columnwidth]{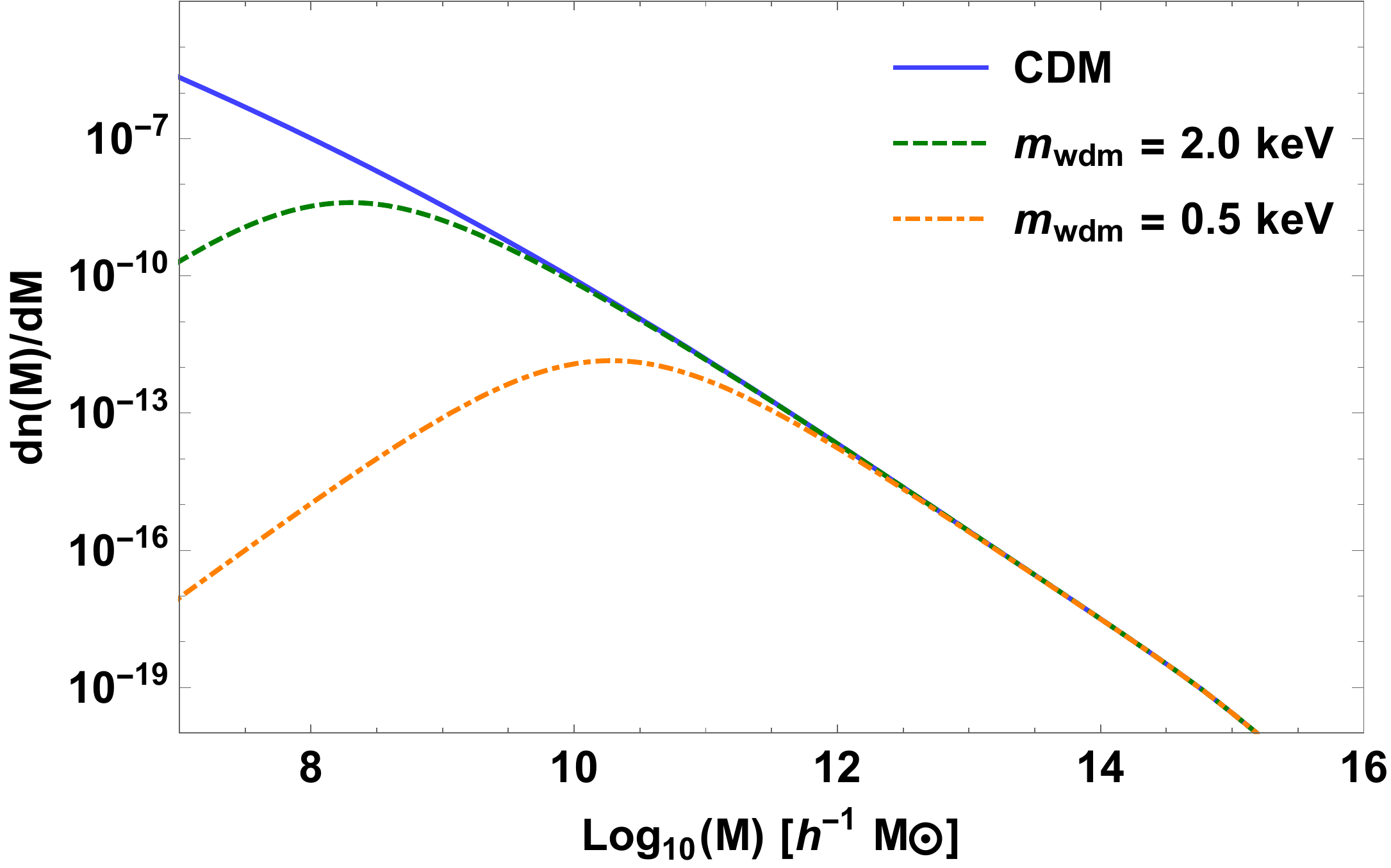}
	\includegraphics[width=\columnwidth]{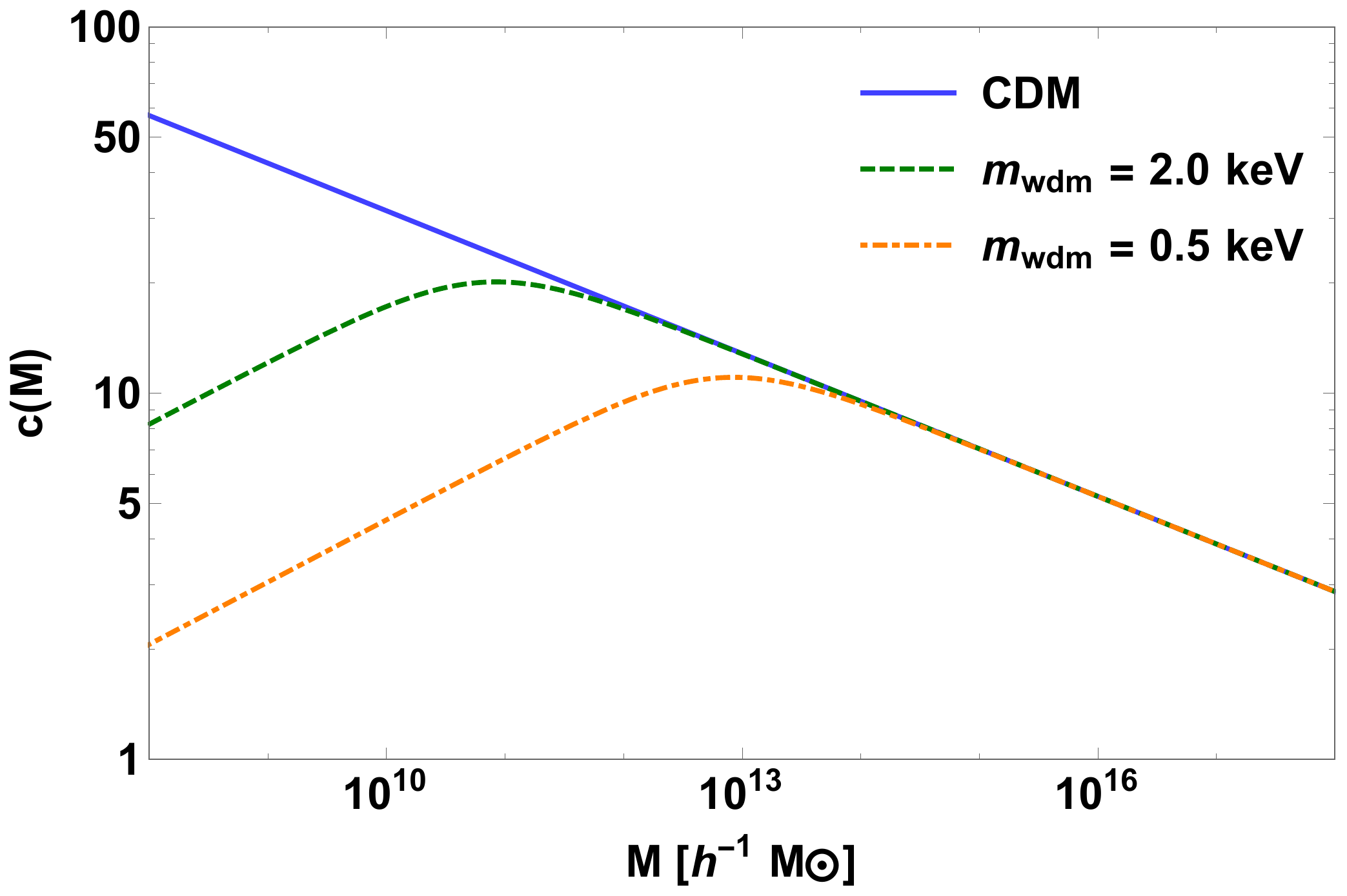}
    \includegraphics[width=\columnwidth]{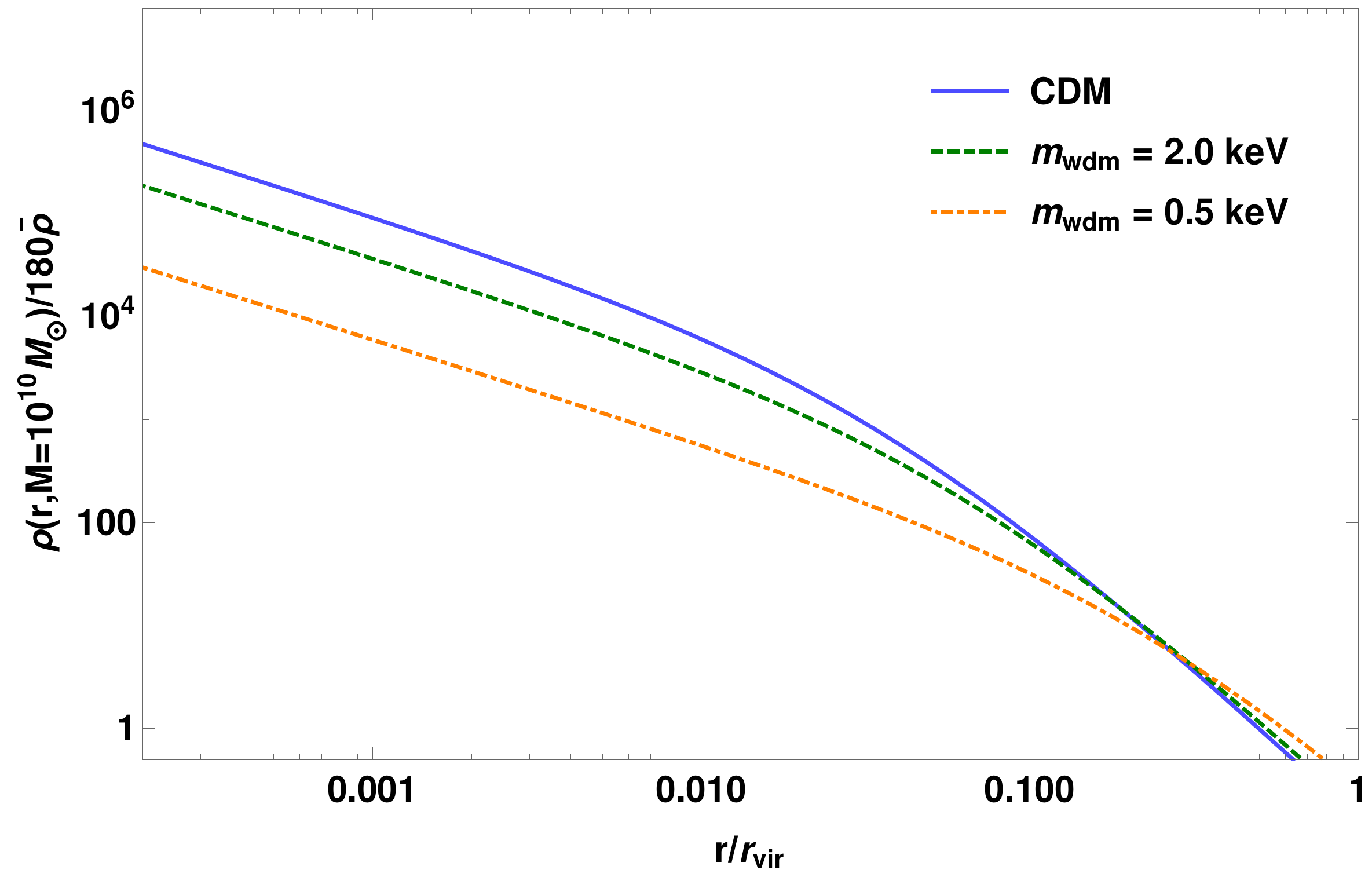}
    
 \includegraphics[width=\columnwidth]{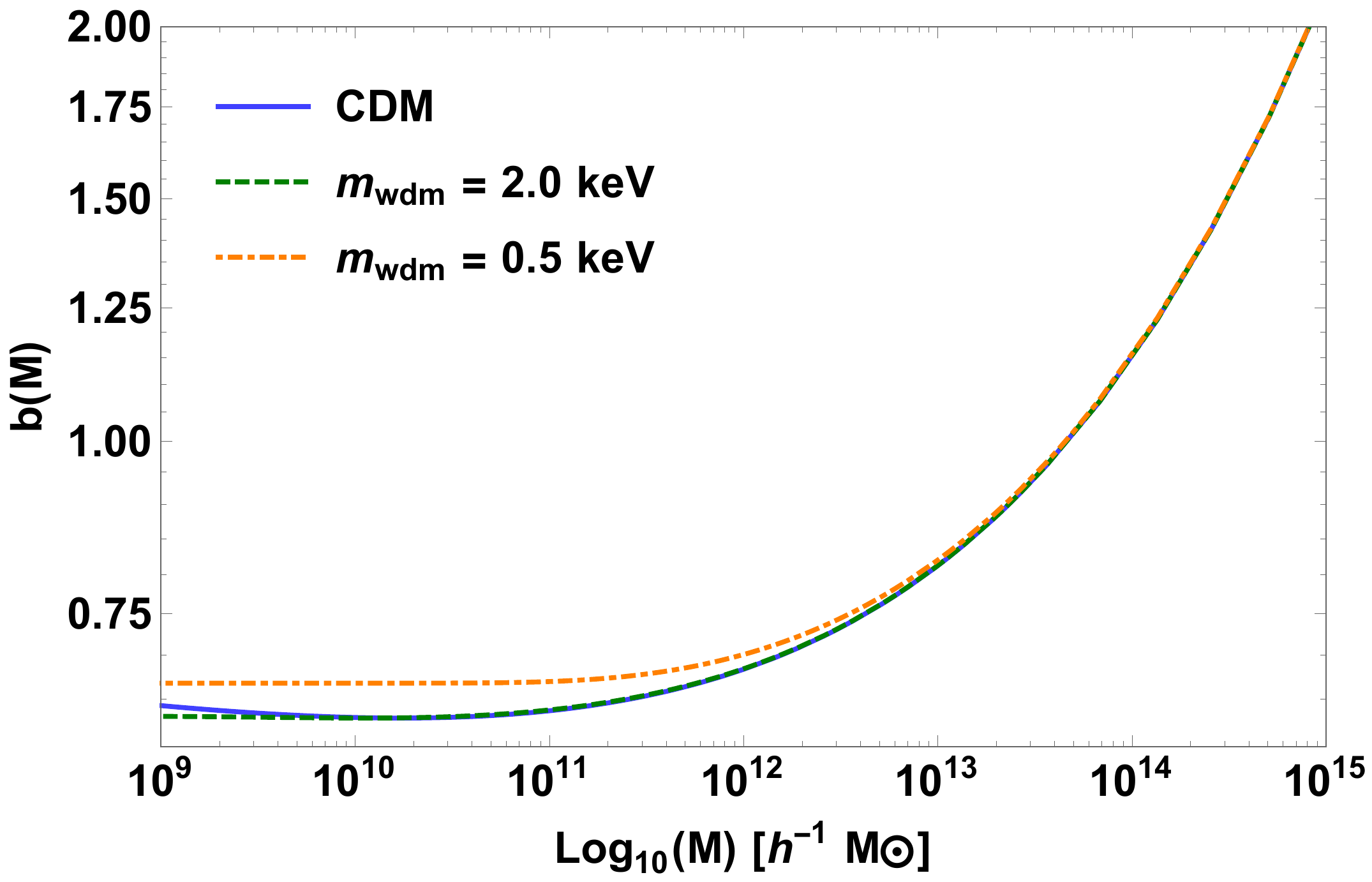}   
	\caption{Jenkins mass function (first), concentration-mass parameter (second), NFW density profile (third) and Tinker halo bias (fourth) for CDM (solid lines), $m_{\text{wdm}}=2.0$\,keV (dashed) and $m_{\text{wdm}}=0.5$\,keV (dot-dashed).}
	\label{fig:nMwdm}
	\end{figure}

For illustration we show the mass function, equation (\ref{eq:fnu}), concentration parameter, equation (\ref{eq:cm}), NFW profile, equation (\ref{eq:nfw}), and halo bias, equation (\ref{eq:bias}) with these modifications for different WDM masses in Fig.~\ref{fig:nMwdm}. As expected, there is a suppression in the number of halos with small masses and it is stronger for lighter WDM particles, and the turnover region is close to the half-mode mass. The concentration inside small halos also gets smoothed in the WDM case, and the inner density of low-mass halos decreases faster for smaller WDM particle masses. This effect is actually the reason for the explanation of the core-cusp problem in WDM. 

\subsection{Non-linear fitting formula for WDM}

Another possible way to deal with non-linear effects is to use fitting formulas. A fitting formula for the non-linear power spectrum of WDM was obtained from simulations. Inspired by the linear fit from (Bode et al. 2001), \citet{viel2012} suggested a formula for the non-linear suppression with an accuracy compared to simulations of $2\%$ at $z<3$ and masses  $m_{\text{wdm}}\geq 0.5$\,keV:
	\begin{align}
	&P^{\text{nonlin}}_{\text{wdm}}(k,z) = P^{\texttt{halofit}}_{\text{cdm}}(k)\left\lbrace 1 + \left[\beta(z) k \right]^{\nu l}  \right\rbrace^{-s/\nu},\\    &\beta(z) = 0.0476 \left( \frac{m_{\text{wdm}}}{1\text{keV}}\right)^{-1.85} \left( \frac{1+z}{2} \right)^{1.3},
	\end{align}
where $\nu = 3,\, l=0.6$ and $s = 0.4$. 

\begin{figure}
	\includegraphics[width=\columnwidth]{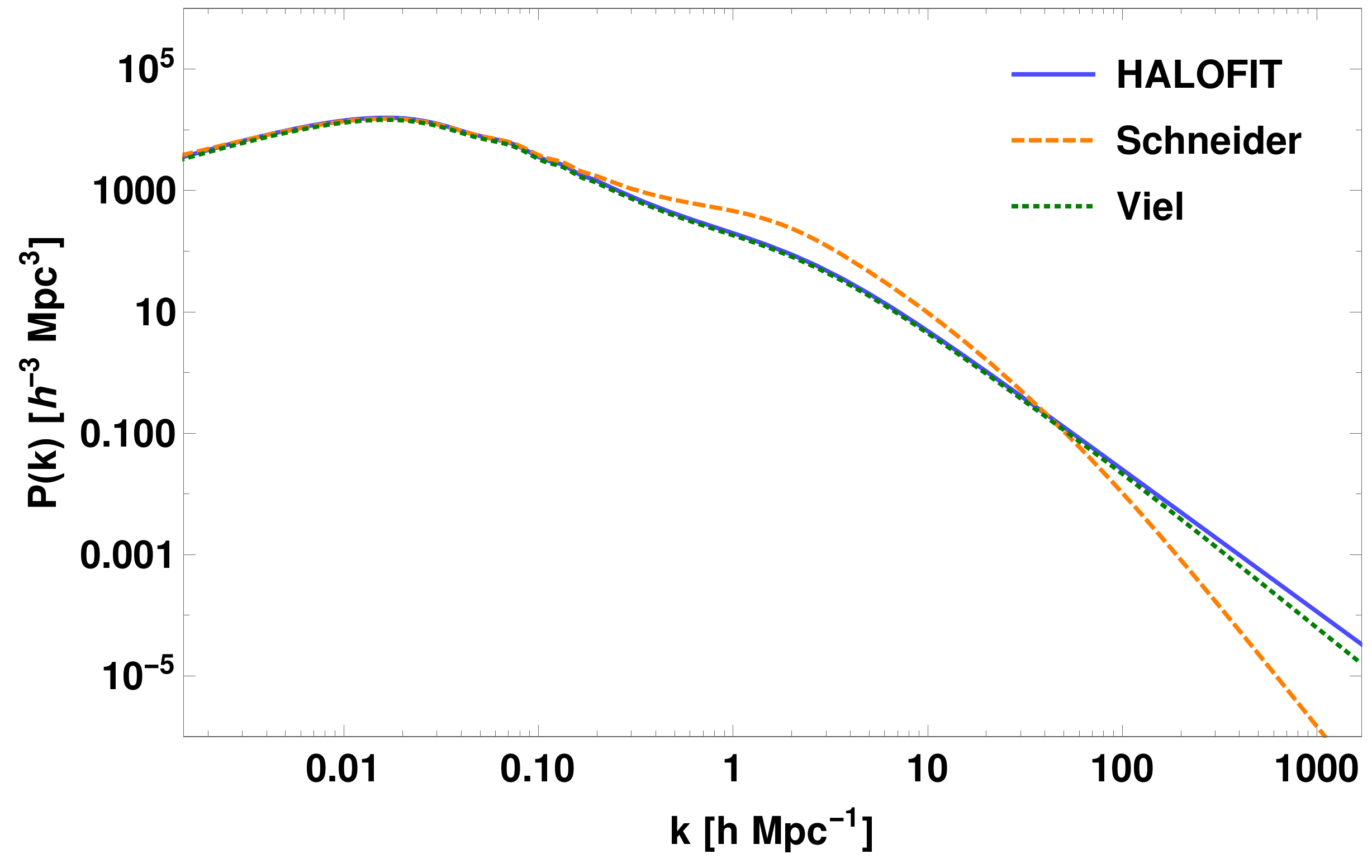}
    \includegraphics[width=\columnwidth]{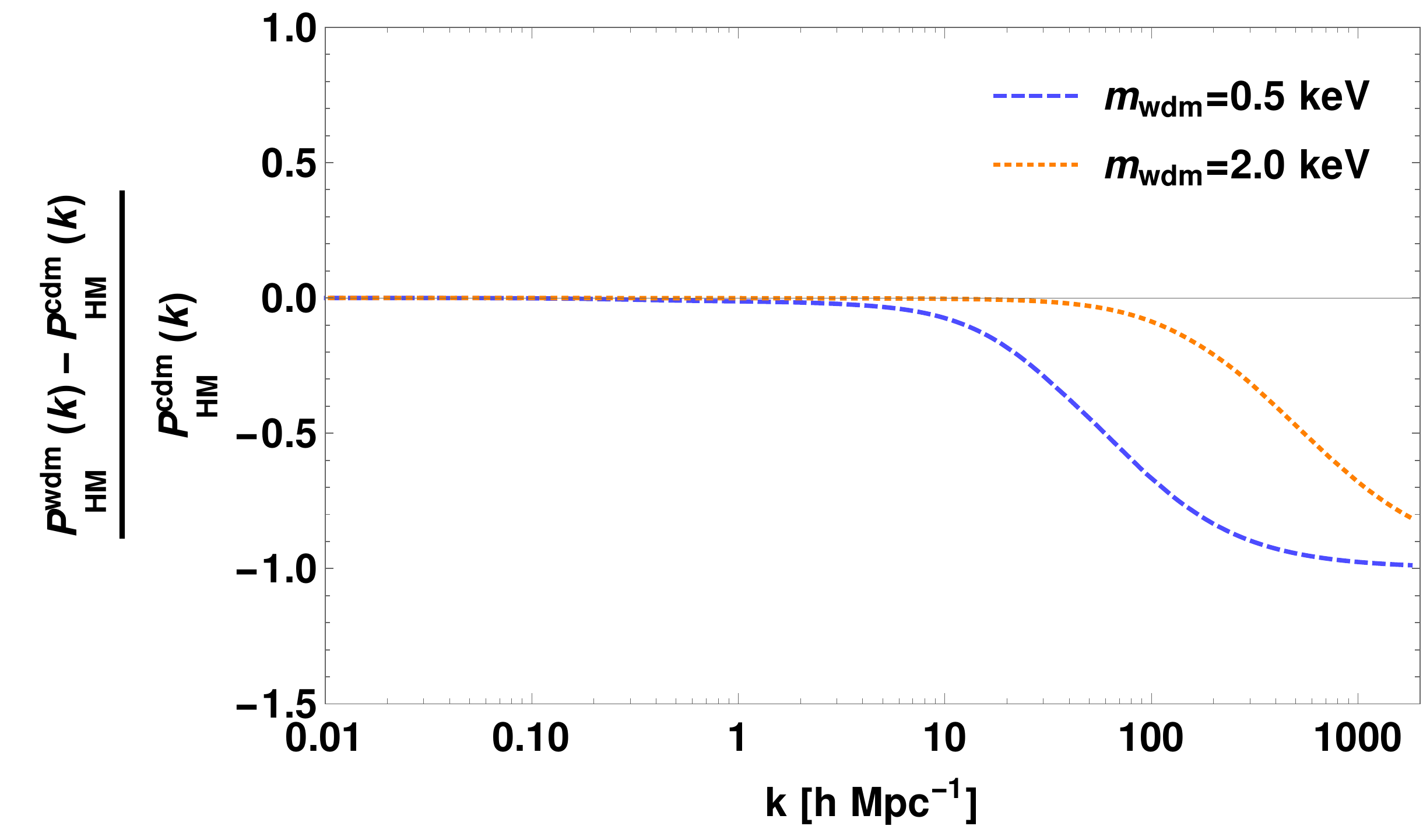}
	\caption{Top: Power spectrum from Viel (dotted) and from modified Halo model (dashed) for $m_{\text{wdm}}=0.5$\,keV, together with CDM \texttt{halofit} (solid line) at z=0.5. Bottom: Difference between non-linear power spectrum from modified halo model for WDM particle masses of $m_{\text{wdm}}=0.5$\,keV (dashed) and $m_{\text{wdm}}=2.0$\,keV (dotted) and CDM at z=0.5.  }
	\label{fig:pkschnvielC}
	\end{figure}

In the Fig.~\ref{fig:pkschnvielC} we show the 3D power spectrum for the modified halo model and the non-linear fitting formula, together with the \texttt{halofit} \citep{halofit} for cold dark matter. Comparing with Fig.~\ref{fig:Plinwdm} one sees that the suppression effect of warm dark matter is much smaller in the non-linear power spectrum than in the linear one. This happens because the linear power spectrum enters in the halo model directly only in the 2-halo term and non-linear effects, which increase the power at small scales, end up by diminishing the WDM imprint. In the bottom panel of same figure we the sensitivity in the suppression of the non-linear power spectrum within the modified halo model for two different values of the WDM mass.   

The halo model predicts a higher suppression of structure at small scales than the non-linear fitting formula from Viel. We opted for an optimistic analysis of the WDM structure and we used the halo model with sharp-k window function to obtain the general constraints for DES and LSST. As we will show in the following section, the use of the Viel non-linear fitting formula results in much weaker constraints for WDM mass.

\section{Forecasting constraints for WDM particle mass from photometric surveys}{\label{sec:Fisher}}

We want to estimate constraints on the sensitivity to the warm dark matter particle mass in a DES-like and a LSST-like surveys, which are wide area photometric surveys. This class of surveys maps galaxies at high redshifts ($z\sim 1$--3) but with poor radial distances resolution, and hence one measures a 2D projection of the galaxy power spectra at different redshift bins.   

\subsection{Angular power spectrum}

In order to write down the angular power spectrum we need first to project the dark matter density field $\delta(\mathbf{x},z)$ along a given direction of the sky using a radial selection function $\phi(z)$, and then expand it in Fourier modes followed by a spherical harmonics decomposition of the plane waves. This leads to the definition of the angular power spectrum $C_\ell$  \citep{dodelson}:
\begin{align}
 \left\langle a_{\ell m}a_{\ell'm}\right\rangle \equiv \delta_{\ell \ell'}\delta_{mm'} C_{\ell},
 \end{align}
 with
	\begin{align}
	a_{\ell m} = 4\pi i^{\ell}\int dz \phi(z)\int \frac{d^{3}k}{(2\pi)^{3}}\delta(\mathbf{k},z)j_{\ell}(kr(z))Y^{*}_{\ell m}(\mathbf{\hat{k}}),
	\end{align}
where $j_{\ell}$ are the spherical Bessel functions of order $\ell$, $\phi(z)$ is the normalised selection function and $r(z)$ is the comoving distance to redshift $z$ given by,
	\begin{align}
	r(z)=\int_{0}^{z}\frac{c}{H(z')}dz',\qquad \frac{H(z)}{H_{0}}=\sqrt{\Omega_{\text{m}}(1+z)^{3}+\Omega_{\Lambda}}
	\end{align}

If the survey is sliced into $n$ redshift bins $i$ the selection function will be given by a sum in each bin:
	\begin{align}
	\phi(z) = \sum_{i}\phi_{i}(z) = \sum_{i}n(z)W_{i}(z),
	\end{align}
where each $\phi_{i}(z)$ is written in terms of the number density of galaxies per unit solid angle and per unit redshift $n(z)$ and a window function 
\begin{align}
W_{i}(z) = \Theta(z-z^{i}_{min})\Theta(z^{i}_{max}-z)
\end{align}
that selects the $i$-th redshift bin. However, in the case of photometric surveys, where there are large uncertainties in redshift measurement, we need to include the probability $P(z^{ph}|z)$ of assigning a true redshift $z$ given a measured photometric redshift $z^{ph}$. The probability function for spectroscopic calibrated galaxies is usually written as a Gaussian distribution \citep{ma2006effects}:
	\begin{align}
	P(z^{ph}|z) = \frac{1}{\sqrt{2\pi}\sigma(z)}\exp\left[-\frac{(z-z^{ph})^{2}}{2\sigma(z)^2} \right],
	\end{align}
and the selection function for a photometric redshift bin $i$ is given by
	\begin{align}
	\phi_{i}(z) = n(z)\int_{z^{i}_{\text{min}}}^{z^{i}_{\text{max}}}dz^{ph}P(z^{ph}|z).
	\end{align}

For DES \citep{sobreira2011,crocce2011modelling} the uncertainty in the photometric redshift is described as $\sigma (z) = 0.03(1+z)$ and the galaxy redshift distribution is parametrized as
	\begin{align}
	n_{\text{DES}}(z) = A \left(\frac{z}{0.5} \right)^{2} \exp{\left( - \frac{z}{0.5} \right)^{1.5}}.
	\end{align}

For LSST $\sigma (z) = 0.05(1+z)$ and $n(z)$ is given by \citep{abell2009lsst} 
\begin{equation}
n_{\text{LSST}}(z) = B z^{2} \exp{\left( -\frac{z}{0.5} \right)}
\end{equation}
where $A$ and $B$ are normalization constants chosen to guarantee that 
	\begin{align}
	\int_{0}^{\infty}dz \phi(z) = N,
	\end{align}
where $N$ is the total number of objects per unit solid angle of the survey. For DES we use $N=15$ arcmin$^{-2}$ \citep{abbott2016dark} and for LSST we take $N=50$ arcmin$^{-2}$ \citep{abell2009lsst}. 

Since we want to analyse highly non-linear scales we are allowed to use Limber approximation \citep{loverde2008extended} to write the angular power spectrum as:
	\begin{align}
	C_{\ell} = \int dz \frac{\phi(z)^{2}}{r(z)^{2}}P_{\text{g}}\left(k=\frac{\ell+1/2}{r(z)},z\right),	
	\end{align} 
where we have introduced the galaxy power spectrum 
	\begin{align}
P_{\text{g}}(k,z)= b^{2}_{\text{g}} \, P_{\text{HM}}(k,z)
	\end{align} 
where $P_{\text{HM}}$ is the halo model 3D matter power spectrum and $b_{\text{g}}$ is the galaxy bias. As we want to give forecasts for a galaxy survey we need to account for the relation between the dark matter and the galaxy distributions, which can be encoded in the galaxy bias. In general this relation can be very complex, but we assume here for simplicity a linear bias model with a redshift-dependent bias.\footnote{The systematic effects of baryonic feedback such as supernova explosions and radiative cooling have been shown to be at the percent level for 
the weak lensing $C_{\ell}$ even at large values of $\ell$ \citep{jing,semboloni,casarini2012tomographic,mead2015accurate} and we expect it to be of the same order for the angular matter power spectrum.}

\begin{figure}
	\includegraphics[width=\columnwidth]{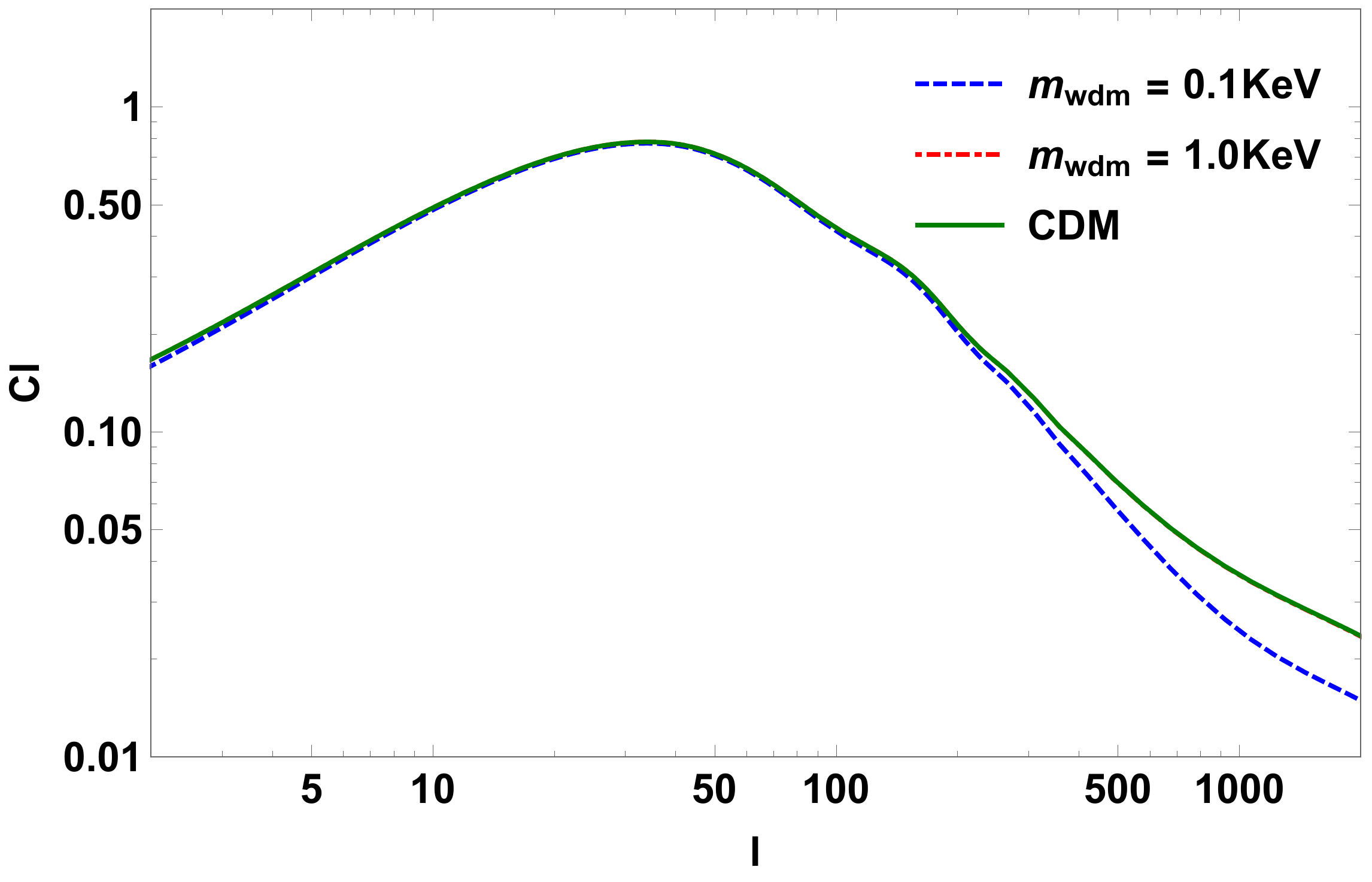}
	\includegraphics[width=\columnwidth]{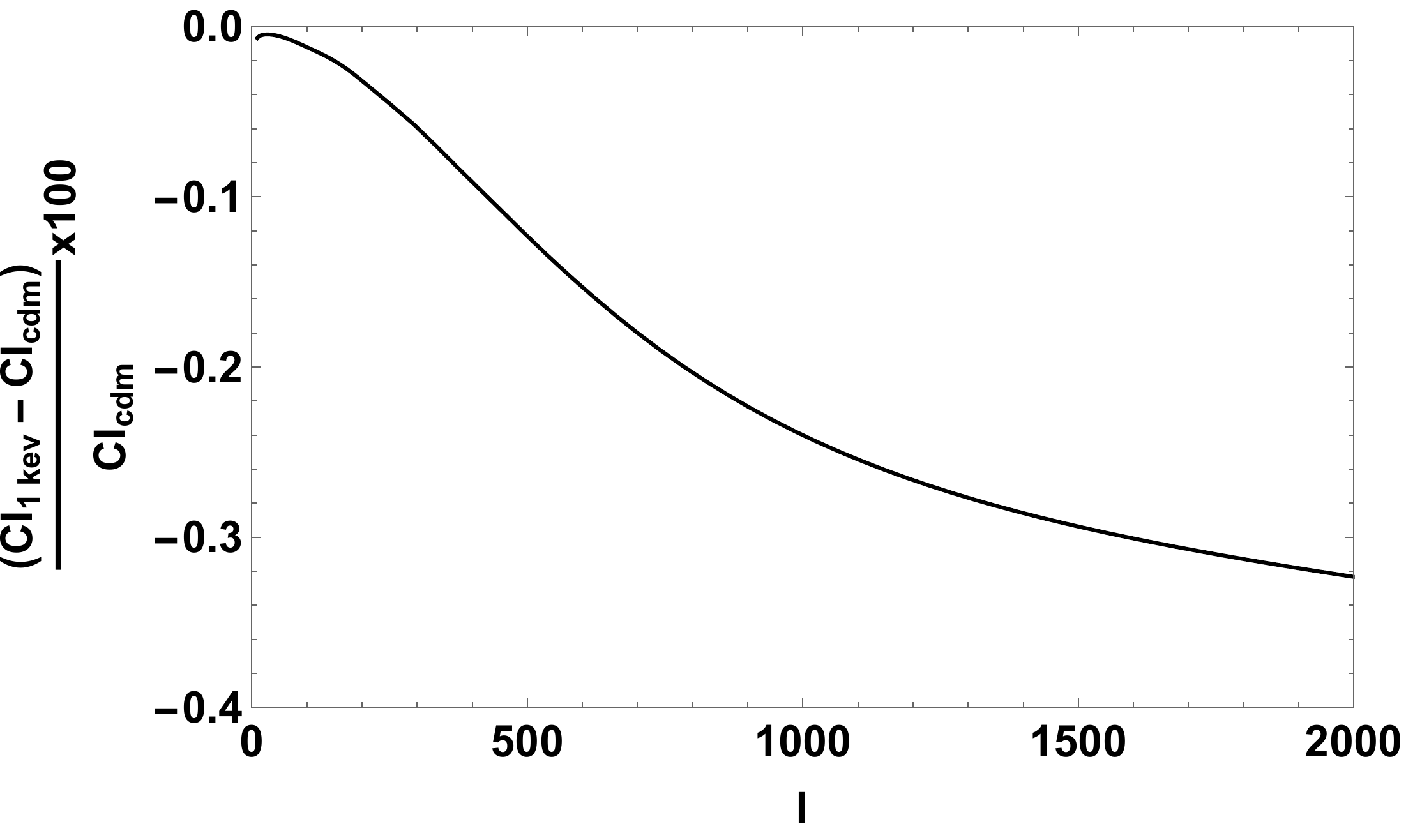}
	\caption{Top: $C_{\ell}$s computed for different WDM particle masses at z=0.5. Bottom: percentage difference between $C_{\ell}$s for $m_{\text{wdm}}=1$\,keV and CDM = $m_{\text{wdm}} = 100$\,MeV.}
	\label{fig:Cls}
	\end{figure}

In Fig.~\ref{fig:Cls} the angular power spectra with LSST selection function at $z=0.5$ for different WDM masses and CDM are shown. In practice we used for the CDM case a $m_{\text{wdm}}=100$\,MeV. The difference between the $C_{\ell}s$ for CDM and a $m_{\text{wdm}}=1$\,keV WDM is about $0.3\%$ at $\ell=2000$ at $z=0$. This difference increases for smaller masses and at higher redshifts where the non-linear effects are less important. In this analysis we will examine both $\ell_{\mbox{max}} = 1000$ and $2000$.

\begin{figure}
	\includegraphics[width=\columnwidth]{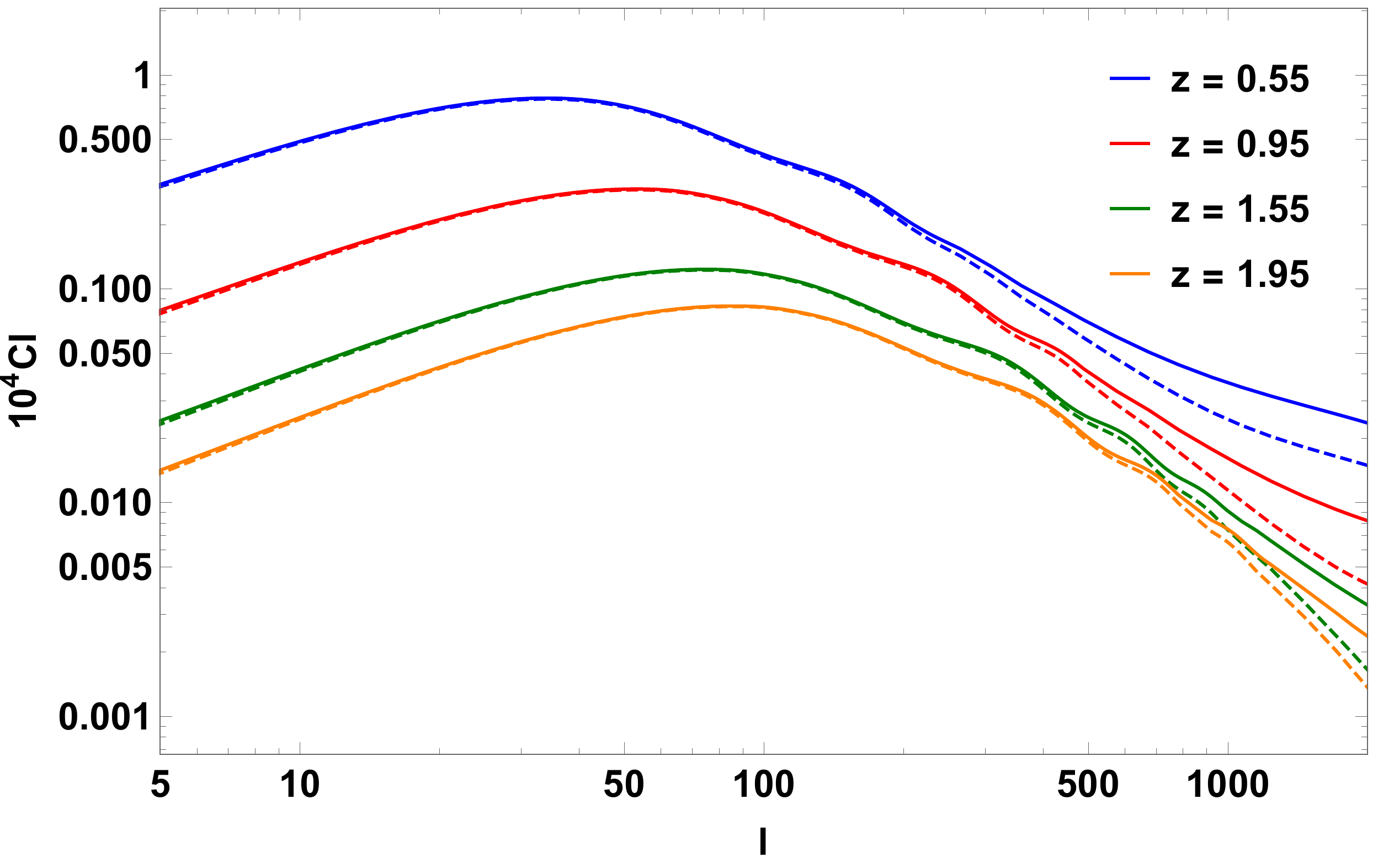}
	\includegraphics[width=\columnwidth]{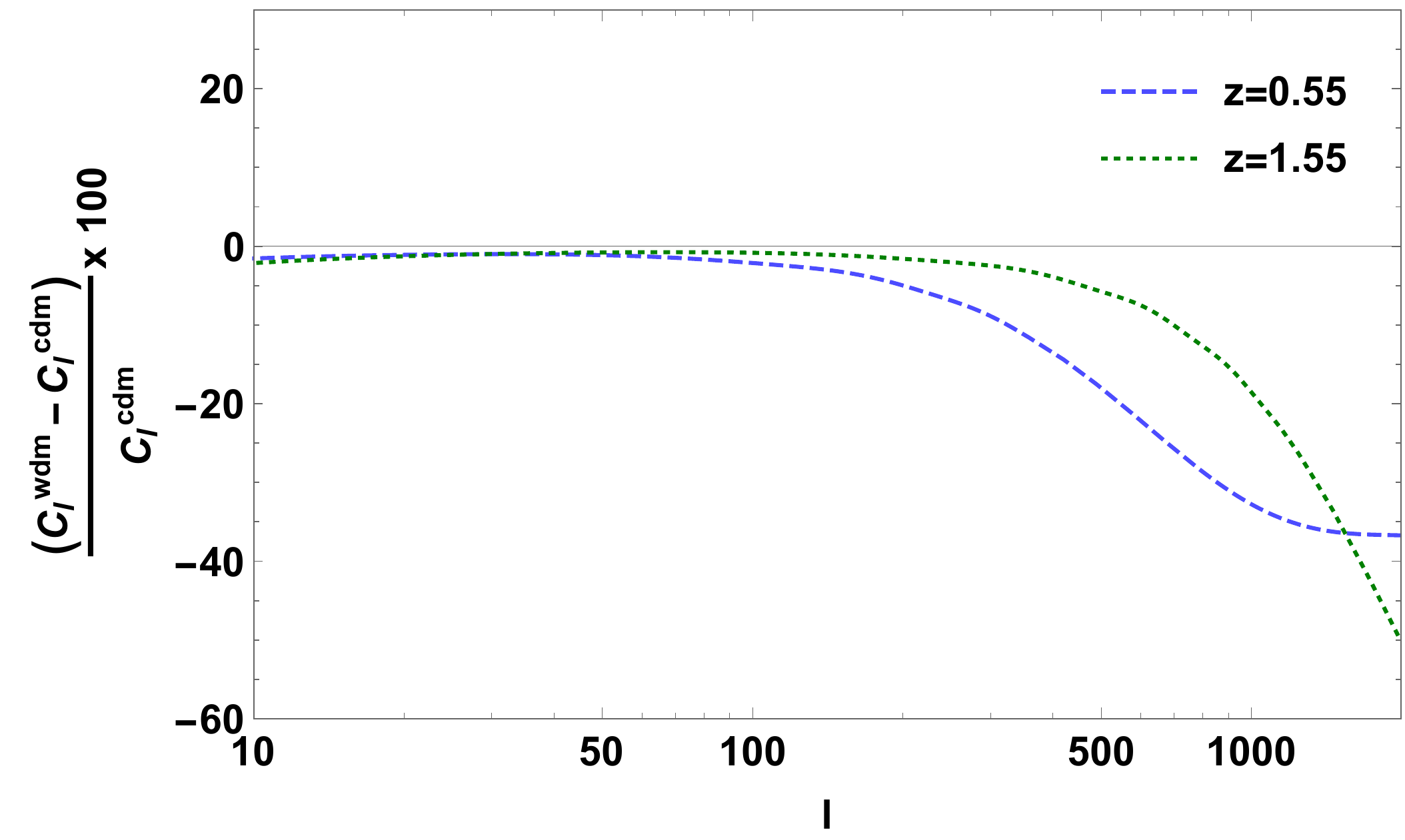}
	\caption{Top: $C_{\ell}$s computed at 4 different redshifts for $m_{\text{wdm}}=0.1$\,keV (Dashed lines) and CDM (Solid lines). Bottom: Percentage difference between $C_{\ell}$s for $m_{\text{wdm}}=0.1$\,keV and CDM at 2 redshifts.}
	\label{fig:Cls4bins}
\end{figure}

In Fig.~\ref{fig:Cls4bins} we show for illustration the resulting angular power spectrum for a LSST-like survey for 4 redshift bins for $m_{\text{wdm}} = 0.1$ KeV compared to the $\Lambda$CDM case. In this case one can see large differences of around $20\%$ even at $\ell=500$ for $z=1.55$.

\subsection{Fisher matrix analysis}

The precision that can be achieved in measurements of cosmological parameters from a given observable is encoded in the Fisher information matrix \citep{amendola2010dark}. 
In the case of the observable being the angular power spectrum the Fisher matrix can be written as  \citep{tegmark1997measuring}
	\begin{align}
	\mathcal{F}_{\alpha \beta}= \sum_{\ell,\ell^\prime}\sum_{i,j}\frac{\partial C_{i}(\ell)}{\partial p_{\alpha}} \left[\left<C_{i}(\ell),C_{j}(\ell^\prime)\right>\right]^{-1} \frac{\partial C_{j}(\ell^\prime)}{\partial p_{\beta}},
	\end{align}
where $p_{\alpha}$ are the parameters of our analysis and $\left<C_{i}(\ell),C_{j}(\ell^\prime)\right>$ is the covariance matrix of the angular power spectrum for redshift bins $i$ and $j$. The estimated 1-$\sigma$ marginalised uncertainty on a parameter $p$ is then:
	\begin{align}
	\sigma_{p}=\sqrt{(\mathcal{F}^{-1})_{pp}}.
	\end{align}

The parameter set chosen is $p_{\alpha} = \left\lbrace m^{-1}_{\text{wdm}},\Omega_{m},b_{g}\right\rbrace$. From the recent DES results \citep{abbott2017dark} we know that the parameters that are most constrained by the 2-point statistics of galaxies and weak lensing are $\Omega_{m}$ and $\sigma_{8}$. Therefore we choose to show our constraints on $m_{\text{wdm}}$ against one of these parameters.

We used four redshift bins for DES forecast between $0.6\leq z\leq 1$ equally spaced with $\Delta z= 0.1$. For LSST we used eight bins also equally spaced with $\Delta z = 0.2$ between $0.4\leq z\leq 2$.   
In the following we will marginalize over the galaxy bias in each redshift bin directly in the Fisher matrix framework assuming a fiducial value of $b_{g}(z) = 1+0.84 \, z$, which is estimated from simulations in \citet{weinbergbias}. 

We assumed that different bins are uncorrelated and that measurements of $C_{\ell}$s are independent, which results in the following covariance matrix for each redshift bin $i$ \citep{hinshaw2003first},
	\begin{align}
	\left<C_{i}(\ell),C_{j}(\ell^\prime)\right> = \frac{1}{f_{\text{sky}}}\frac{2}{2l+1}C_{i}(\ell)^2 \delta_{\ell \ell^\prime}\delta_{ij},
	\end{align}
where we have used the $f_{\text{sky}}$ approximation. For DES we adopt $f_{\text{sky}}= 1/8$ and for LSST $f_{\text{sky}}= 0.485$, according to the area intended to be mapped by the surveys.

	\begin{figure*}
	\centering
	\includegraphics[width=\columnwidth]{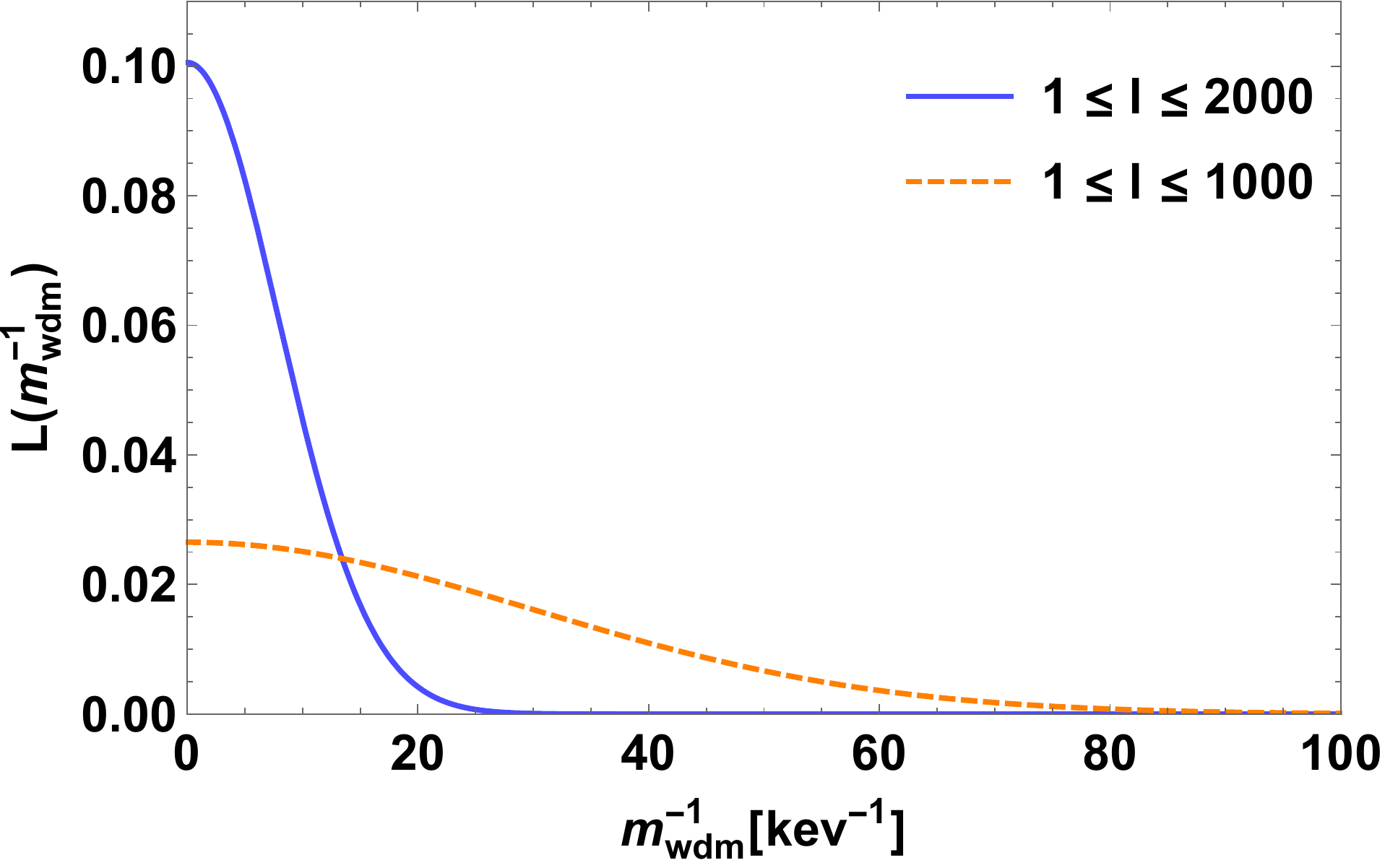}
	\includegraphics[width=\columnwidth]{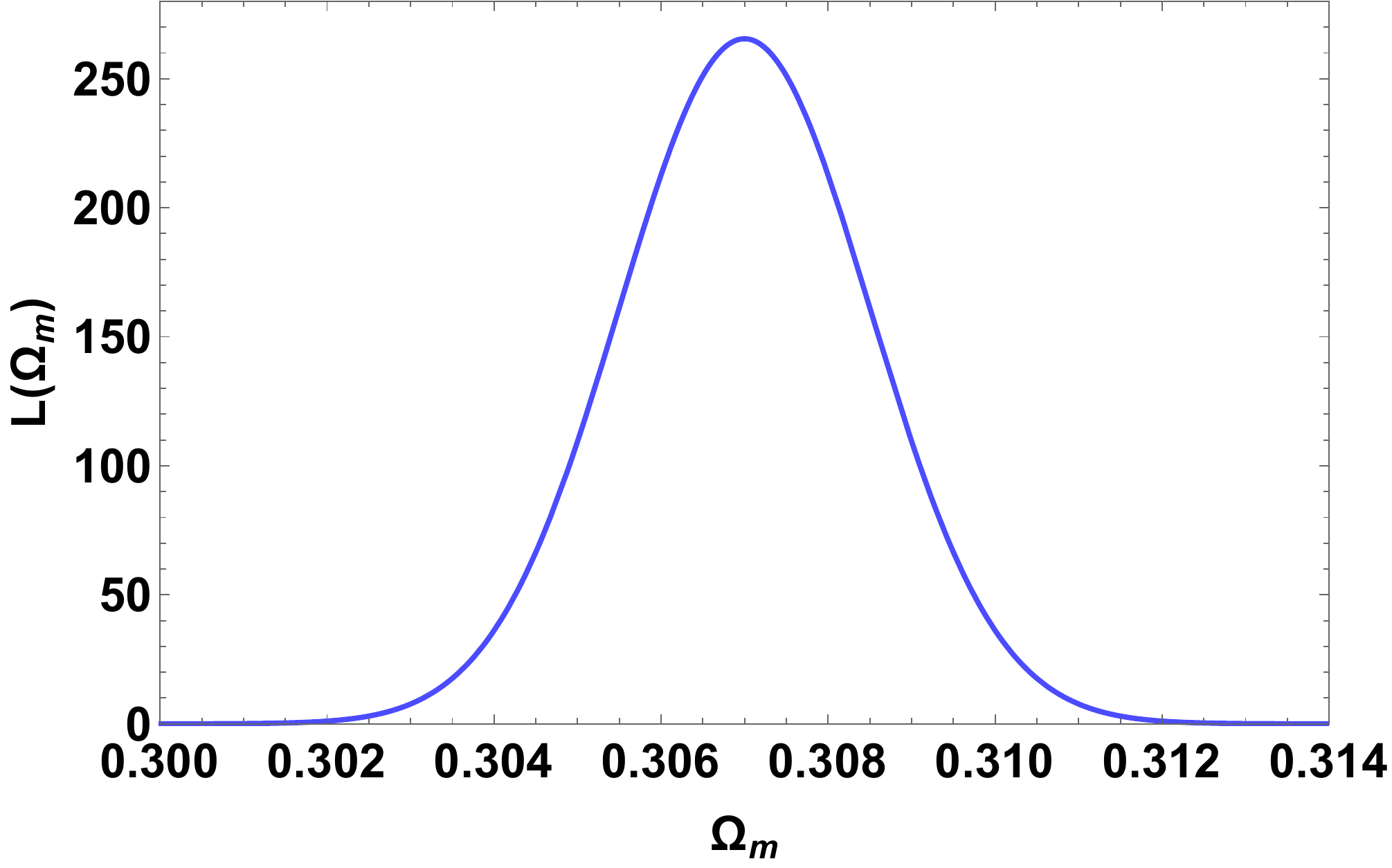}
	\includegraphics[width=\columnwidth]{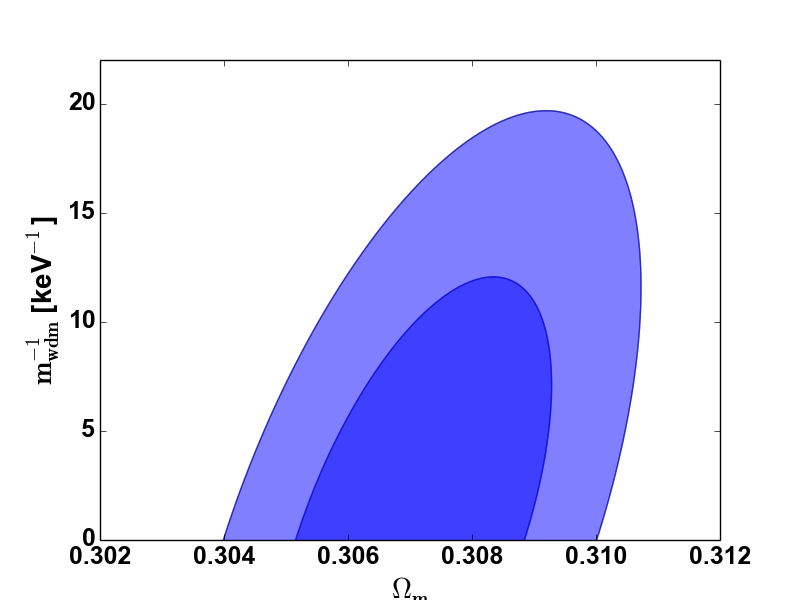}
	\caption{Top left: normalized likelihoods of $m^{-1}_{\text{wdm}}$ marginalized over $\Omega_{\text{m}}$ and $b_{\text{g}}$. The dashed line shows the probability function for calculations done with $l$ until 1000. We see that decreasing the non-linear regime in the computations has a great impact in error estimation (see Section \ref{sec:discussion}). Top right: likelihood of $\Omega_{\text{m}}$ with $m^{-1}_{\text{wdm}}$ and $b_{\text{g}}$ marginalized. Bottom: the expected error ellipsis for $\Omega_{\text{m}}$ and $m^{-1}_{\text{wdm}}$ with $b_{\text{g}}$ marginalized. The light blue and dark blue curves represent a $2\sigma$ and $1\sigma$ confidence region, respectively. All plots are DES forecasts.}
	\label{fig:likeOmmxDES}
	\end{figure*}
    
    \begin{figure*}
	\centering
	\includegraphics[width=\columnwidth]{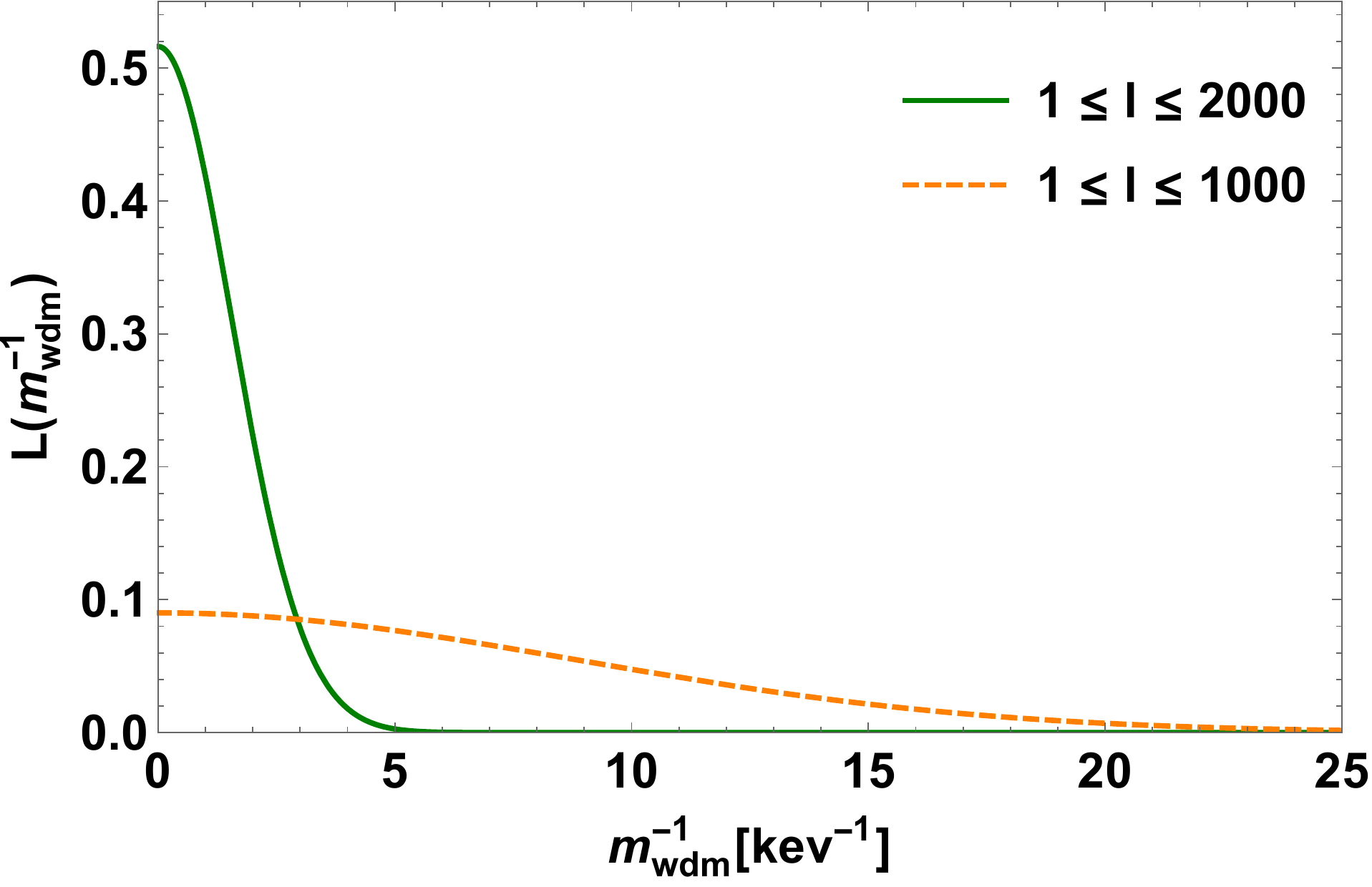}
	\includegraphics[width=\columnwidth]{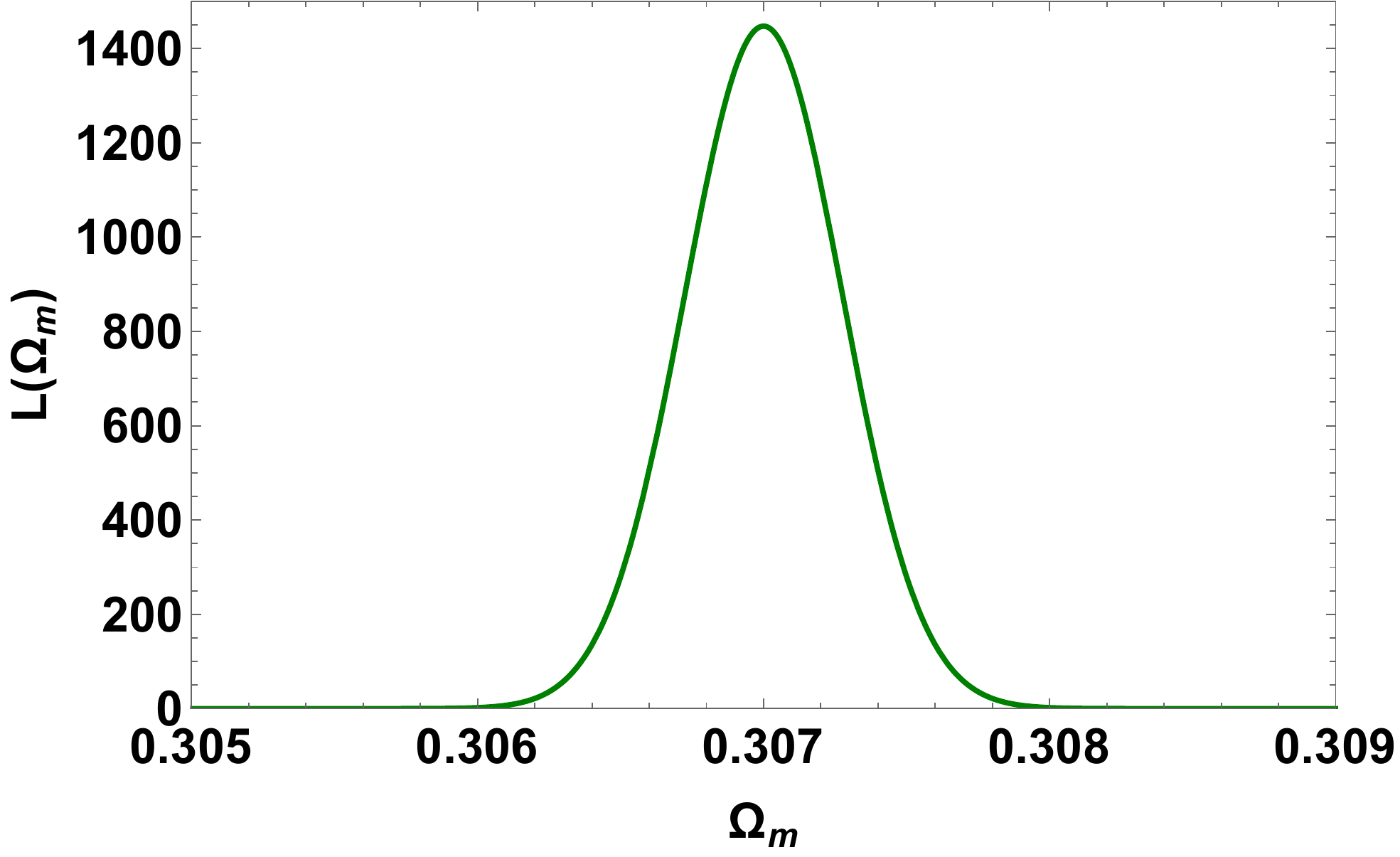}
	\includegraphics[width=\columnwidth]{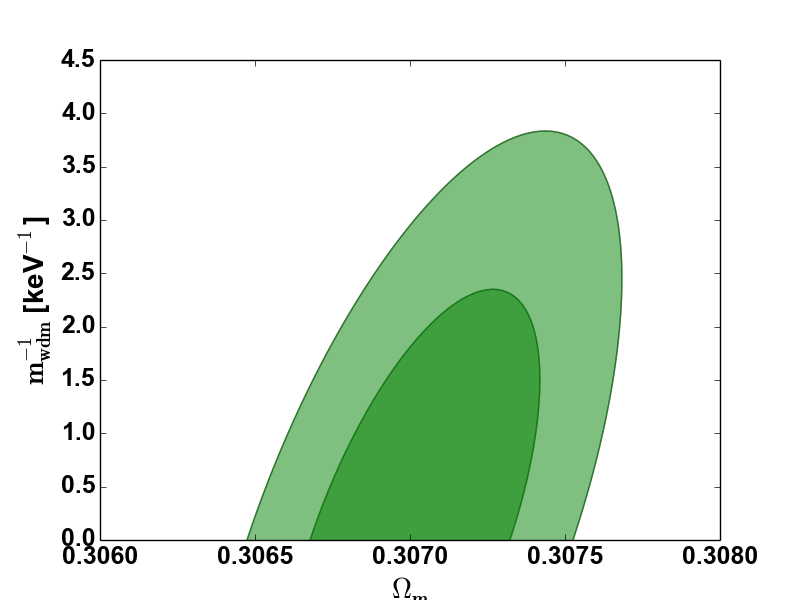}
	\caption{Top left: normalized likelihoods of $m^{-1}_{\text{wdm}}$ marginalized over $\Omega_{\text{m}}$ and $b_{\text{g}}$. Top right: likelihood of $\Omega_{\text{m}}$ with $m^{-1}_{\text{wdm}}$ and $b_{\text{g}}$ marginalized. Bottom: the expected error ellipse for $\Omega_{\text{m}}$ and $m^{-1}_{\text{wdm}}$ with $b_{\text{g}}$ marginalized. The light green and dark green curves represent a $2\sigma$ and $1\sigma$ confidence region, respectively. All plots are LSST forecasts.}
	\label{fig:likeOmmxLSST}
	\end{figure*}
    
    \begin{figure*}
	\centering
	\includegraphics[width=\columnwidth]{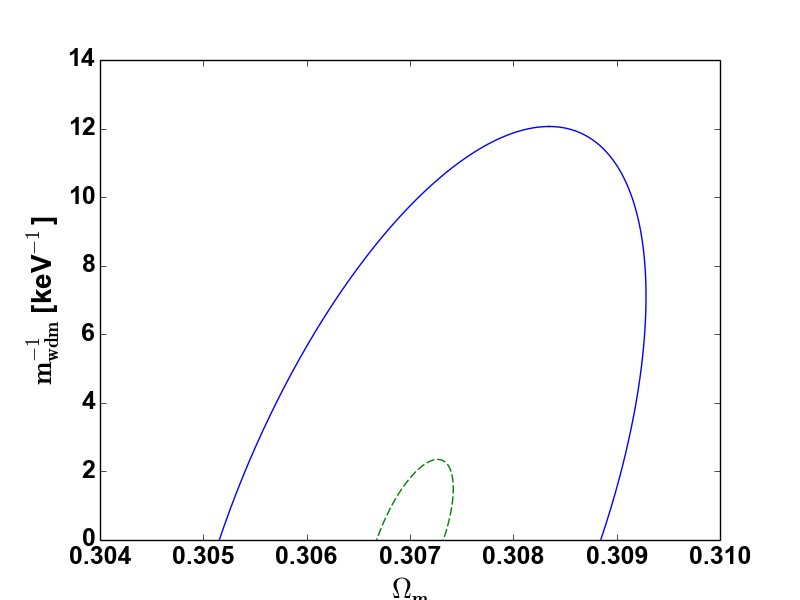}
    \includegraphics[width=\columnwidth]{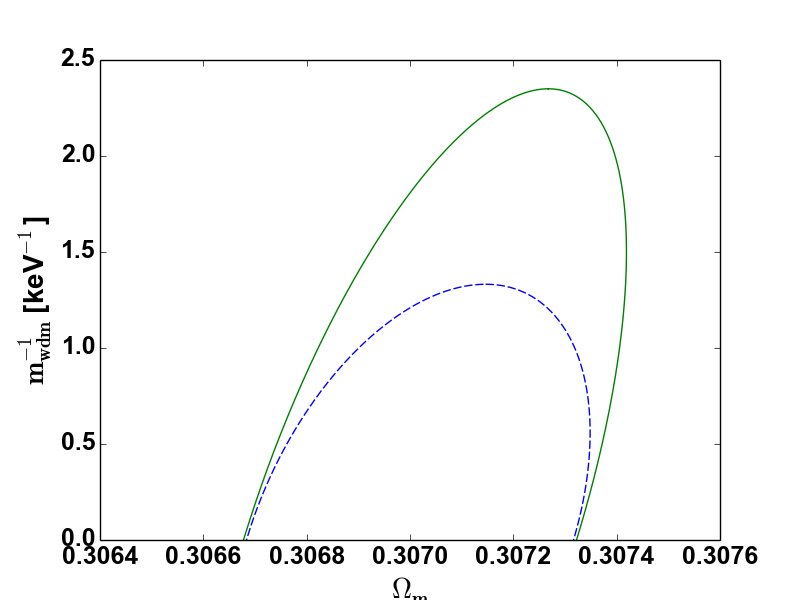}
	\caption{Left: Comparison between DES (blue line) and LSST (green dashed) $1\sigma$ error ellipse. Right: 1$\sigma$ error ellipse for LSST (green line) and combined result with shear power spectra from EUCLID (blue dashed).}
	\label{fig:deslsst}
	\end{figure*}

The parameters which we aim to constrain are $\Omega_{\text{m}}$ and $m^{-1}_{\text{wdm}}$. We chose to work with $m^{-1}_{\text{wdm}}$ instead of $m_{\text{wdm}}$ to recover our fiducial $\Lambda$CDM model in the limit where the parameter related to the mass of dark matter particle goes to zero rather than infinity. 
This leads to numerical complications when calculating derivatives over $m^{-1}_{\text{wdm}}$ (as well for $m_{\text{wdm}}$) as the $C_{\ell}$s become insensitive to small variations close to the fiducial value for the warm dark matter mass. This insensitivity implies that small variations in the mass parameter induce variations in the $C_{\ell}$s that are smaller than numerical noise. 
We handled this subtlety by calculating the numerical derivative around the fiducial model of an interpolation function constructed with $C_{\ell}$s for various different WDM particle masses at each redshift bin. 

For our fiducial $\Lambda$CDM model we used $\Omega_{\text{m}} = 0.307,\, \Omega_{\text{b}} = 0.048,\, \Omega_{\Lambda}=0.693,\, n_{\text{s}}=0.968,\, w=-1,\, h=0.679$ following Planck results \citep{planck} and $m_{\text{cdm}} = 100$\,MeV. The parameters constraints are shown in Fig.~\ref{fig:likeOmmxDES} (for DES-like surveys) and in Fig.~\ref{fig:likeOmmxLSST} (for LSST-like surveys). 

For DES, at $1\sigma$ confidence level we found a precision of 0.49\% in the measurement of $\Omega_{\text{m}}$ ($\sigma\left( \Omega_{\text{m}} \right)= 0.0015$). One obtains a sensitivity to an upper limit on the inverse of the WDM mass, which translates into a lower limit in $m_{\text{wdm}}$. In this case we obtain $m_{\text{wdm}}>126$\,eV 
 at $1\sigma$.

For LSST, at $1\sigma$ confidence level we found a precision of 0.09\% in the measurement of $\Omega_{\text{m}}$ ($\sigma \left( \Omega_{\text{m}}\right) = 0.0003$). 
For the WDM particle mass the lower limit found was $m_{\text{wdm}}>647$\,eV  also at $1\sigma$. Fig.~\ref{fig:deslsst} shows the $1\sigma$ error ellipse for DES and LSST for comparison.  

    \begin{figure*}
	\centering
	\includegraphics[width=\columnwidth]{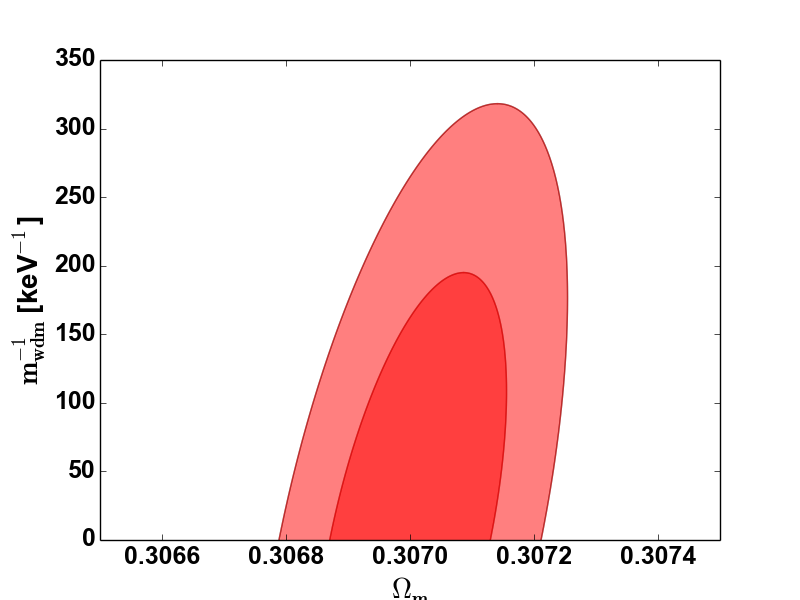}
	\caption{1$\sigma$ (dark red) and 2$\sigma$ (light red) error ellipse for LSST using Viel non-linear fitting formula.}
	\label{fig:erviel}
	\end{figure*}

As pointed out in the end of Section~\ref{sec:freewdm} and can be seen from Fig.~\ref{fig:erviel}, we get poor constraints for WDM particle mass when using the non-linear dark matter power spectrum of Viel et al. (2012). At $1\sigma$ we could place a lower limit of $m_{wdm}>7.8\,$eV for the LSST. This justifies why we called the use of the Halo Model an optimistic approach, as otherwise using a best fit from simulations gives barely no constraints on the mass.

\subsection{Combined result with weak lensing}

It is interesting to compare our results with a Fisher matrix forecast for WDM from cosmic shear power spectrum \citep{markovic2011constraining}. An estimated lower bound of $m_{\text{wdm}}>935$~eV is obtained from an Euclid-like weak lensing survey where they use multipoles as large as $\ell = 10^4$. This is comparable to our estimate from a LSST-like survey. We should also notice that there is no dependence on the galaxy bias in this case and we assumed a diagonal covariance matrix from weak lensing for simplicity.

The results for the combined $1\sigma$ error ellipse for LSST and EUCLID are shown in the right panel of Fig.~\ref{fig:deslsst}. For the combined analysis we could place a lower limit of $m_{\text{wdm}}>1.14$\,keV at $1\sigma$ and a precision of $0.07\%$ in the measurement of $\Omega_{m}$ ($\sigma \left( \Omega_{\text{m}}\right) = 0.0002$).

\section{Discussion}{\label{sec:discussion}}

In this work we made the first estimate of constraints on the WDM particle mass using the galaxy angular power spectrum for a DES and a LSST-like photometric surveys. We used a well-know parametrization of the modified linear power spectrum and a modified halo model with a sharp-k window function and a new concentration-mass parameter based on N-body simulations of warm dark matter for the non-linear power spectrum. 

We estimated a lower bound of $m_{\text{wdm}}>126$~eV for DES and $m_{\text{wdm}}>647$~eV for LSST at $1 \sigma$ confidence level on the particle mass using the angular power spectrum.  

It is interesting to compare our results with a Fisher matrix forecast for WDM from cosmic shear power spectrum \citep{markovic2011constraining}. An estimated lower bound of $m_{\text{wdm}}>645$~eV is obtained from an Euclid-like weak lensing survey where they use multipoles as large as $\ell = 10^4$. This is comparable to our estimate from a LSST-like survey. We should also notice that there is no dependence on the galaxy bias in this case. For the combined probe we found a lower limit of $m_{\text{wdm}}>1.14$\,keV.

Our results degrade rapidly if we leave out very small scales from the analysis. This is expected, since as we showed above the main differences in the power spectrum appear at small scales or large redshifts. We also have results for $\ell <1000$ shown in the upper left panel of Fig.~\ref{fig:likeOmmxDES} and Fig.~\ref{fig:likeOmmxLSST}. In this case the bounds on the mass are reduced to $m_{\text{wdm}}>33$~eV for DES and $m_{\text{wdm}}>113$~eV for LSST.

We should recall that there are other ways to modify the halo model to account for WDM. One method worth mentioning is (Schneider et al. 2012), where instead of imposing the normalization condition in equation~(\ref{eq:norm}), one could add another term in the statistics to represent the fraction of dark matter that didn't collapse into halos due to the free-streaming of WDM particles. Then, there would be a correlation function between the dark matter inside and outside halos, which has to be taken into account.

In addition to modifying the halo model, it would also be interesting to consider the halo occupation distribution model (HOD). As in practice we observe galaxies of baryonic matter instead of dark matter, it is relevant to work with a model for the occupation of objects inside halos. This would improve on our naive linear bias model. It would be as well of great importance to correctly include baryonic effects on structure formation, once this scenario is fully understood with the help of simulations.

Our estimated bounds are not competitive with bounds from Ly-$\alpha$ mentioned in the introduction but we think they should be explored anyways with real data and afterwards used in combinations of different probes, including CMB. 

\section*{Acknowledgements}

JSM thanks the financial support from CAPES. RR is partially supported by a Fapesp grant and by the Instituto Nacional de Ci\^encia e Tecnologia (INCT) e-Universe project (CNPq grant 465376/2014-2). We thank useful discussions with Alexander Mead.



\bibliographystyle{mnras}
\bibliography{ref}

\begin{thebibliography}{}
\makeatletter
\relax
\def\mn@urlcharsother{\let\do\@makeother \do\$\do\&\do\#\do\^\do\_\do\%\do\~}
\def\mn@doi{\begingroup\mn@urlcharsother \@ifnextchar [ {\mn@doi@}
  {\mn@doi@[]}}
\def\mn@doi@[#1]#2{\def\@tempa{#1}\ifx\@tempa\@empty \href
  {http://dx.doi.org/#2} {doi:#2}\else \href {http://dx.doi.org/#2} {#1}\fi
  \endgroup}
\def\mn@eprint#1#2{\mn@eprint@#1:#2::\@nil}
\def\mn@eprint@arXiv#1{\href {http://arxiv.org/abs/#1} {{\tt arXiv:#1}}}
\def\mn@eprint@dblp#1{\href {http://dblp.uni-trier.de/rec/bibtex/#1.xml}
  {dblp:#1}}
\def\mn@eprint@#1:#2:#3:#4\@nil{\def\@tempa {#1}\def\@tempb {#2}\def\@tempc
  {#3}\ifx \@tempc \@empty \let \@tempc \@tempb \let \@tempb \@tempa \fi \ifx
  \@tempb \@empty \def\@tempb {arXiv}\fi \@ifundefined
  {mn@eprint@\@tempb}{\@tempb:\@tempc}{\expandafter \expandafter \csname
  mn@eprint@\@tempb\endcsname \expandafter{\@tempc}}}

\bibitem[\protect\citeauthoryear{Abbott et~al.,}{Abbott
  et~al.}{2016}]{abbott2016dark}
Abbott T.,  et~al., 2016, Monthly Notices of the Royal Astronomical Society,
  460, 1270

\bibitem[\protect\citeauthoryear{{Abbott} et~al.,}{{Abbott}
  et~al.}{2017a}]{2017arXiv170801530D}
{Abbott} et~al., 2017a, preprint, \href
  {http://adsabs.harvard.edu/abs/2017arXiv170801530D} {} (\mn@eprint {arXiv}
  {1708.01530})

\bibitem[\protect\citeauthoryear{Abbott et~al.,}{Abbott
  et~al.}{2017b}]{abbott2017dark}
Abbott T.,  et~al., 2017b, arXiv preprint arXiv:1708.01530

\bibitem[\protect\citeauthoryear{Abell et~al.,}{Abell
  et~al.}{2009}]{abell2009lsst}
Abell P.~A.,  et~al., 2009, Technical report, Lsst science book, version 2.0

\bibitem[\protect\citeauthoryear{Ade et~al.,}{Ade et~al.}{2016}]{planck}
Ade P.~A.,  et~al., 2016, Astronomy and Astrophysics, 594

\bibitem[\protect\citeauthoryear{Amendola \& Tsujikawa}{Amendola \&
  Tsujikawa}{2010}]{amendola2010dark}
Amendola L.,  Tsujikawa S.,  2010, Dark energy: theory and observations.
Cambridge University Press

\bibitem[\protect\citeauthoryear{Bardeen, Bond, Kaiser  \& Szalay}{Bardeen
  et~al.}{1986}]{bardeen1986statistics}
Bardeen J.~M.,  Bond J.,  Kaiser N.,   Szalay A.,  1986, The Astrophysical
  Journal, 304, 15

\bibitem[\protect\citeauthoryear{Benson et~al.,}{Benson
  et~al.}{2012}]{benson2012dark}
Benson A.~J.,  et~al., 2012, Monthly Notices of the Royal Astronomical Society,
  428, 1774

\bibitem[\protect\citeauthoryear{Bode, Ostriker  \& Turok}{Bode
  et~al.}{2001}]{bode}
Bode P.,  Ostriker J.~P.,   Turok N.,  2001, The Astrophysical Journal, 556, 93

\bibitem[\protect\citeauthoryear{Bullock}{Bullock}{2013}]{bullock2013notes}
Bullock J.~S.,  2013, Local Group Cosmology, 20, 95

\bibitem[\protect\citeauthoryear{Casarini, Bonometto, Borgani, Dolag, Murante,
  Mezzetti, Tornatore  \& La~Vacca}{Casarini
  et~al.}{2012}]{casarini2012tomographic}
Casarini L.,  Bonometto S.~A.,  Borgani S.,  Dolag K.,  Murante G.,  Mezzetti
  M.,  Tornatore L.,   La~Vacca G.,  2012, Astronomy \& Astrophysics, 542, A126

\bibitem[\protect\citeauthoryear{Col{\'\i}n, Valenzuela  \&
  Avila-Reese}{Col{\'\i}n et~al.}{2008}]{colin2008structure}
Col{\'\i}n P.,  Valenzuela O.,   Avila-Reese V.,  2008, The Astrophysical
  Journal, 673, 203

\bibitem[\protect\citeauthoryear{Cooray \& Sheth}{Cooray \&
  Sheth}{2002}]{sheth}
Cooray A.,  Sheth R.,  2002, Physics Reports, 372, 1

\bibitem[\protect\citeauthoryear{Crocce, Cabr{\'e}  \& Gazta{\~n}aga}{Crocce
  et~al.}{2011}]{crocce2011modelling}
Crocce M.,  Cabr{\'e} A.,   Gazta{\~n}aga E.,  2011, Monthly Notices of the
  Royal Astronomical Society, 414, 329

\bibitem[\protect\citeauthoryear{De~Blok}{De~Blok}{2010}]{de2010core}
De~Blok W.,  2010, Advances in Astronomy, 2010

\bibitem[\protect\citeauthoryear{Diemand, Moore  \& Stadel}{Diemand
  et~al.}{2004}]{diemand}
Diemand J.,  Moore B.,   Stadel J.,  2004, Monthly Notices of the Royal
  Astronomical Society, 353, 624

\bibitem[\protect\citeauthoryear{Dodelson}{Dodelson}{2003}]{dodelson}
Dodelson S.,  2003, Modern cosmology.
Academic press

\bibitem[\protect\citeauthoryear{Dunstan, Abazajian, Polisensky  \&
  Ricotti}{Dunstan et~al.}{2011}]{dunstan2011halo}
Dunstan R.~M.,  Abazajian K.~N.,  Polisensky E.,   Ricotti M.,  2011, arXiv
  preprint arXiv:1109.6291

\bibitem[\protect\citeauthoryear{{Fattahi}, {Navarro}, {Sawala}, {Frenk},
  {Sales}, {Oman}, {Schaller}  \& {Wang}}{{Fattahi}
  et~al.}{2016}]{2016arXiv160706479F}
{Fattahi} A.,  {Navarro} J.~F.,  {Sawala} T.,  {Frenk} C.~S.,  {Sales} L.~V.,
  {Oman} K.,  {Schaller} M.,   {Wang} J.,  2016, preprint, \href
  {http://adsabs.harvard.edu/abs/2016arXiv160706479F} {} (\mn@eprint {arXiv}
  {1607.06479})

\bibitem[\protect\citeauthoryear{Garrison-Kimmel, Boylan-Kolchin, Bullock  \&
  Kirby}{Garrison-Kimmel et~al.}{2014}]{garrison2014too}
Garrison-Kimmel S.,  Boylan-Kolchin M.,  Bullock J.~S.,   Kirby E.~N.,  2014,
  Monthly Notices of the Royal Astronomical Society, 444, 222

\bibitem[\protect\citeauthoryear{Hernquist}{Hernquist}{1990}]{hernquist}
Hernquist L.,  1990, The Astrophysical Journal, 356, 359

\bibitem[\protect\citeauthoryear{Hinshaw et~al.,}{Hinshaw
  et~al.}{2003}]{hinshaw2003first}
Hinshaw G.,  et~al., 2003, The Astrophysical Journal Supplement Series, 148,
  135

\bibitem[\protect\citeauthoryear{Ir{\v{s}}i{\v{c}} et~al.,}{Ir{\v{s}}i{\v{c}}
  et~al.}{2017}]{lyalphamx}
Ir{\v{s}}i{\v{c}} V.,  et~al., 2017, arXiv preprint arXiv:1702.01764

\bibitem[\protect\citeauthoryear{Jenkins, Frenk, White, Colberg, Cole, Evrard,
  Couchman  \& Yoshida}{Jenkins et~al.}{2001}]{pssimulations}
Jenkins A.,  Frenk C.,  White S.~D.,  Colberg J.,  Cole S.,  Evrard A.~E.,
  Couchman H.,   Yoshida N.,  2001, Monthly Notices of the Royal Astronomical
  Society, 321, 372

\bibitem[\protect\citeauthoryear{Klypin, Kravtsov, Valenzuela  \& Prada}{Klypin
  et~al.}{1999}]{klypin1999missing}
Klypin A.,  Kravtsov A.~V.,  Valenzuela O.,   Prada F.,  1999, The
  Astrophysical Journal, 522, 82

\bibitem[\protect\citeauthoryear{Kolb \& Turner}{Kolb \& Turner}{1990}]{kolb}
Kolb E.~W.,  Turner M.~S.,  1990, Front. Phys.,Vol. 69,, 1

\bibitem[\protect\citeauthoryear{{Leo}, {Baugh}, {Li}  \& {Pascoli}}{{Leo}
  et~al.}{2017}]{2017arXiv171202742L}
{Leo} M.,  {Baugh} C.~M.,  {Li} B.,   {Pascoli} S.,  2017, preprint, \href
  {http://adsabs.harvard.edu/abs/2017arXiv171202742L} {} (\mn@eprint {arXiv}
  {1712.02742})

\bibitem[\protect\citeauthoryear{Lin, Jing, Mao, Gao  \& McCarthy}{Lin
  et~al.}{2006}]{jing}
Lin W.-P.,  Jing Y.,  Mao S.,  Gao L.,   McCarthy I.,  2006, The Astrophysical
  Journal, 651, 636

\bibitem[\protect\citeauthoryear{LoVerde \& Afshordi}{LoVerde \&
  Afshordi}{2008}]{loverde2008extended}
LoVerde M.,  Afshordi N.,  2008, Physical Review D, 78, 123506

\bibitem[\protect\citeauthoryear{{Lovell}, {Gonzalez-Perez}, {Bose},
  {Boyarsky}, {Cole}, {Frenk}  \& {Ruchayskiy}}{{Lovell}
  et~al.}{2017}]{2017MNRAS.468.2836L}
{Lovell} M.~R.,  {Gonzalez-Perez} V.,  {Bose} S.,  {Boyarsky} A.,  {Cole} S.,
  {Frenk} C.~S.,   {Ruchayskiy} O.,  2017, \mn@doi [\mnras]
  {10.1093/mnras/stx621}, \href
  {http://adsabs.harvard.edu/abs/2017MNRAS.468.2836L} {468, 2836}

\bibitem[\protect\citeauthoryear{Ma, Hu  \& Huterer}{Ma
  et~al.}{2006}]{ma2006effects}
Ma Z.,  Hu W.,   Huterer D.,  2006, The Astrophysical Journal, 636, 21

\bibitem[\protect\citeauthoryear{Markovic, Bridle, Slosar  \& Weller}{Markovic
  et~al.}{2011}]{markovic2011constraining}
Markovic K.,  Bridle S.,  Slosar A.,   Weller J.,  2011, Journal of Cosmology
  and Astroparticle Physics, 2011, 022

\bibitem[\protect\citeauthoryear{Marsh}{Marsh}{2016}]{fuzzy}
Marsh D.~J.,  2016, arXiv preprint arXiv:1605.05973

\bibitem[\protect\citeauthoryear{Massara, Villaescusa-Navarro  \& Viel}{Massara
  et~al.}{2014}]{massara2014halo}
Massara E.,  Villaescusa-Navarro F.,   Viel M.,  2014, Journal of Cosmology and
  Astroparticle Physics, 2014, 053

\bibitem[\protect\citeauthoryear{Mead, Peacock, Heymans, Joudaki  \&
  Heavens}{Mead et~al.}{2015}]{mead2015accurate}
Mead A.,  Peacock J.,  Heymans C.,  Joudaki S.,   Heavens A.,  2015, Monthly
  Notices of the Royal Astronomical Society, 454, 1958

\bibitem[\protect\citeauthoryear{Menci, Fiore  \& Lamastra}{Menci
  et~al.}{2012}]{menci2012galaxy}
Menci N.,  Fiore F.,   Lamastra A.,  2012, Monthly Notices of the Royal
  Astronomical Society, 421, 2384

\bibitem[\protect\citeauthoryear{Mo \& White}{Mo \&
  White}{1996}]{mo1996analytic}
Mo H.,  White S.~D.,  1996, Monthly Notices of the Royal Astronomical Society,
  282, 347

\bibitem[\protect\citeauthoryear{Moore, Ghigna, Governato, Lake, Quinn, Stadel
  \& Tozzi}{Moore et~al.}{1999}]{moore}
Moore B.,  Ghigna S.,  Governato F.,  Lake G.,  Quinn T.,  Stadel J.,   Tozzi
  P.,  1999, The Astrophysical Journal Letters, 524, L19

\bibitem[\protect\citeauthoryear{Navarro \& White}{Navarro \&
  White}{1996}]{nfw}
Navarro J.,  White S.~D.,  1996, in SYMPOSIUM-INTERNATIONAL ASTRONOMICAL UNION.
  pp 255--258

\bibitem[\protect\citeauthoryear{Navarro, Frenk  \& White}{Navarro
  et~al.}{1997}]{navarro}
Navarro J.~F.,  Frenk C.~S.,   White S.~D.,  1997, The Astrophysical Journal,
  490, 493

\bibitem[\protect\citeauthoryear{Oh, De~Blok, Brinks, Walter  \&
  Kennicutt~Jr}{Oh et~al.}{2011}]{oh2011dark}
Oh S.-H.,  De~Blok W.,  Brinks E.,  Walter F.,   Kennicutt~Jr R.~C.,  2011, The
  Astronomical Journal, 141, 193

\bibitem[\protect\citeauthoryear{Peebles \& Nusser}{Peebles \&
  Nusser}{2010}]{peebles2010nearby}
Peebles P.,  Nusser A.,  2010, Nature, 465, 565

\bibitem[\protect\citeauthoryear{Press \& Schechter}{Press \&
  Schechter}{1974}]{pressschechter}
Press W.~H.,  Schechter P.,  1974, The Astrophysical Journal, 187, 425

\bibitem[\protect\citeauthoryear{Salucci, Lapi, Tonini, Gentile, Yegorova  \&
  Klein}{Salucci et~al.}{2007}]{salucci}
Salucci P.,  Lapi A.,  Tonini C.,  Gentile G.,  Yegorova I.,   Klein U.,  2007,
  Monthly Notices of the Royal Astronomical Society, 378, 41

\bibitem[\protect\citeauthoryear{Schneider}{Schneider}{2014}]{schneider}
Schneider A.,  2014, arXiv preprint arXiv:1412.2133

\bibitem[\protect\citeauthoryear{Schneider, Smith, Macci{\`o}  \&
  Moore}{Schneider et~al.}{2012}]{schneider2012}
Schneider A.,  Smith R.~E.,  Macci{\`o} A.~V.,   Moore B.,  2012, Monthly
  Notices of the Royal Astronomical Society, 424, 684

\bibitem[\protect\citeauthoryear{Semboloni, Hoekstra, Schaye, van Daalen  \&
  McCarthy}{Semboloni et~al.}{2011}]{semboloni}
Semboloni E.,  Hoekstra H.,  Schaye J.,  van Daalen M.~P.,   McCarthy I.~G.,
  2011, Monthly Notices of the Royal Astronomical Society, 417, 2020

\bibitem[\protect\citeauthoryear{Sheth \& Tormen}{Sheth \&
  Tormen}{1999}]{sheth1999large}
Sheth R.~K.,  Tormen G.,  1999, Monthly Notices of the Royal Astronomical
  Society, 308, 119

\bibitem[\protect\citeauthoryear{Sheth, Mo  \& Tormen}{Sheth
  et~al.}{2001}]{shethellipsoidal}
Sheth R.~K.,  Mo H.,   Tormen G.,  2001, Monthly Notices of the Royal
  Astronomical Society, 323, 1

\bibitem[\protect\citeauthoryear{Smith \& Markovic}{Smith \&
  Markovic}{2011}]{smithtesting}
Smith R.~E.,  Markovic K.,  2011, Physical Review D, 84, 063507

\bibitem[\protect\citeauthoryear{Sobreira, de Simoni, Rosenfeld, da Costa, Maia
   \& Makler}{Sobreira et~al.}{2011}]{sobreira2011}
Sobreira F.,  de Simoni F.,  Rosenfeld R.,  da Costa L.,  Maia M.,   Makler M.,
   2011, Physical Review D, 84, 103001

\bibitem[\protect\citeauthoryear{Swaters, Sancisi, Van~Albada  \& Van
  Der~Hulst}{Swaters et~al.}{2009}]{swaters}
Swaters R.,  Sancisi R.,  Van~Albada T.,   Van Der~Hulst J.,  2009, Astronomy
  \& Astrophysics, 493, 871

\bibitem[\protect\citeauthoryear{Takahashi, Sato, Nishimichi, Taruya  \&
  Oguri}{Takahashi et~al.}{2012}]{halofit}
Takahashi R.,  Sato M.,  Nishimichi T.,  Taruya A.,   Oguri M.,  2012, The
  Astrophysical Journal, 761, 152

\bibitem[\protect\citeauthoryear{Tegmark}{Tegmark}{1997}]{tegmark1997measuring}
Tegmark M.,  1997, Physical Review Letters, 79, 3806

\bibitem[\protect\citeauthoryear{Tikhonov \& Klypin}{Tikhonov \&
  Klypin}{2009}]{tikhonov2009emptiness}
Tikhonov A.~V.,  Klypin A.,  2009, Monthly Notices of the Royal Astronomical
  Society, 395, 1915

\bibitem[\protect\citeauthoryear{Tinker, Kravtsov, Klypin, Abazajian, Warren,
  Yepes, Gottl{\"o}ber  \& Holz}{Tinker et~al.}{2008}]{tinker2008toward}
Tinker J.,  Kravtsov A.~V.,  Klypin A.,  Abazajian K.,  Warren M.,  Yepes G.,
  Gottl{\"o}ber S.,   Holz D.~E.,  2008, The Astrophysical Journal, 688, 709

\bibitem[\protect\citeauthoryear{Tinker, Robertson, Kravtsov, Klypin, Warren,
  Yepes  \& Gottl{\"o}ber}{Tinker et~al.}{2010}]{tinker2010large}
Tinker J.~L.,  Robertson B.~E.,  Kravtsov A.~V.,  Klypin A.,  Warren M.~S.,
  Yepes G.,   Gottl{\"o}ber S.,  2010, The Astrophysical Journal, 724, 878

\bibitem[\protect\citeauthoryear{Viel, Lesgourgues, Haehnelt, Matarrese  \&
  Riotto}{Viel et~al.}{2005}]{vielwdm}
Viel M.,  Lesgourgues J.,  Haehnelt M.~G.,  Matarrese S.,   Riotto A.,  2005,
  Physical Review D, 71, 063534

\bibitem[\protect\citeauthoryear{Viel, Markovi{\v{c}}, Baldi  \& Weller}{Viel
  et~al.}{2012}]{viel2012}
Viel M.,  Markovi{\v{c}} K.,  Baldi M.,   Weller J.,  2012, Monthly Notices of
  the Royal Astronomical Society, 421, 50

\bibitem[\protect\citeauthoryear{Weinberg, Dav{\'e}, Katz  \&
  Hernquist}{Weinberg et~al.}{2004}]{weinbergbias}
Weinberg D.~H.,  Dav{\'e} R.,  Katz N.,   Hernquist L.,  2004, The
  Astrophysical Journal, 601, 1

\bibitem[\protect\citeauthoryear{White}{White}{2001}]{white2001redshift}
White M.,  2001, Monthly Notices of the Royal Astronomical Society, 321, 1

\bibitem[\protect\citeauthoryear{Zentner}{Zentner}{2005}]{zentner2005halo}
Zentner A.,  2005, The Halo Model

\makeatother
\end{thebibliography}




\bsp	
\label{lastpage}
\end{document}